\documentclass{aa}
\usepackage{graphicx}
\usepackage{color}
\usepackage{txfonts}
\usepackage{natbib}

\def\sfr{\mbox{SFR}}
\def\ltir{L_{\rm TIR}}
\def\lfuv{L_{\rm FUV}}
\def\lsf{L_{\rm SF}}
\def\lakari{L_{\rm AKARI}}
\begin{document}
   \title{Star formation and dust extinction properties of local galaxies from AKARI-GALEX All-Sky Surveys:}
   \subtitle{First results from most secure multiband sample from FUV to FIR}
   
   \author{T.\ T.\ Takeuchi\inst{1}
				\and
				V.\ Buat\inst{2}
				\and
				S.\ Heinis\inst{3}
				\and
				E.\ Giovannoli\inst{2}
				\and
				F.-T.\ Yuan\inst{4}
				\and
				J.\ Iglesias-P\'{a}ramo\inst{5}
				\and
				K.\ L.\ Murata\inst{4}
				\and	
				D.\ Burgarella\inst{2}}

   \offprints{T.\ T.\ Takeuchi}

   \institute{Institute for Advanced Research, Nagoya University, Furo-cho, Chikusa-ku, Nagoya 464--8601, JAPAN\\
				\email{takeuchi@iar.nagoya-u.ac.jp}
				\and
				Laboratoire d'Astrophysique de Marseille, OAMP, Universit\'e Aix-Marseille, 
				CNRS, 38 rue Fr\'ed\'eric Joliot-Curie, 13388 Marseille cedex 13, FRANCE\\
				\email{veronique.buat, elodie.giovannoli, denis.burgarella@oamp.fr}
				\and
				Department of Physics \& Astronomy, The Johns Hopkins University, 
				3701 San Martin Drive, Baltimore, MD 21218, USA\\
				\email{sebastien@pha.jhu.edu}
				\and
				Division of Particle and Astrophysical Sciences,  
				Nagoya University, Furo-cho, Chikusa-ku, Nagoya 464--8602, JAPAN\\
				\email{yuan.fangting, murata.katsuhiro@g.mbox.nagoya-u.ac.jp}
				\and
				Instituto de Astrof\'{\i}sica de Andaluc\'{\i}a (IAA, CSIC), Camino Bajo de Hu\'{e}tor
				50, 18008 Granada, SPAIN\\
				\email{jiglesia@iaa.es}
				}

   \date{}

% \abstract{}{}{}{}{} 
% 5 {} token are mandatory
 
  \abstract
  % context heading (optional)  $z$ 
   {}
   {We explore spectral energy distributions (SEDs), star formation,
     and dust extinction properties of galaxies in the Local
     Universe.}
   {The AKARI All-Sky Survey provided the first bright point source
     catalog detected at $90\;\mu$m.  Starting from this catalog, we
     selected galaxies by matching AKARI sources with those in the
     IRAS PSC$z$.  
	 Next, we have measured total GALEX FUV and NUV flux
     densities by a photometry software we have
     specifically developed for this purpose.  
	 Then, we have matched this sample with SDSS and 2MASS
     galaxies. 
	 By this procedure, we obtained the basic sample which
     consists of 776 galaxies.  
	 After removing objects with photometry contaminated by foreground 
	 sources (mainly in SDSS), we have defined the ``secure sample'' which
     contains 607 galaxies. 
	 Using this galaxy sample, we have
     explored various properties of galaxies related to star
     formation and dust extinction.  
	 }
     {{
	 The sample galaxies have redshifts $\la 0.15$}, and
     their 90-$\mu$m luminosities range from $10^6$ to $10^{12}\;L_\odot$, 
	 with a peak at $10^{10}\;L_\odot$.  
	 {
	 The SEDs display a large variety, especially more than four
     orders of magnitude at M-FIR
	 }, but if we sort the sample with respect to $90\;\mu$m, their average SED 
	 shows a coherent trend: 
	 the more luminous at $90\;\mu$m, the redder the global SED becomes.  
	 The $M_r$-$\mbox{NUV}-r$ color-magnitude relation of our sample 
	 does not show bimodality, and the distribution is centered on the green
     valley between the blue cloud and red sequence seen in optical surveys. 
	 We have established formulae to
     convert FIR luminosity from AKARI bands to the total infrared
     (IR) luminosity $\ltir$.  
	 With these formulae, we calculated
     the star formation directly visible with FUV and hidden by dust.  
	 The luminosity related to star formation activity
     ($\lsf$) is dominated by $\ltir$ even if we take into account
     the far-infrared (FIR) emission from dust heated by old stars.
	 At high star formation rate (SFR) ($> 20\;\mbox{M}_\odot\,\mbox{yr}^{-1}$), 
	 the fraction of directly visible SFR, $\sfr_{\rm FUV}$, decreases.
	 We also estimated the FUV attenuation $A_{\rm FUV}$ from FUV-to-total
     IR (TIR) luminosity ratio.  We also examined the
     $\ltir/\lfuv$-UV slope ($\mbox{FUV}- \mbox{NUV}$) relation.
     The majority of the sample has $\ltir/\lfuv$
     ratios 5 to 10 times lower than expected from the local
     starburst relation, while some LIRGs and all the ULIRGs of
     this sample have higher $\ltir/\lfuv$ ratios.
	 We found that the attenuation indicator
     $\ltir/\lfuv$ is correlated to the stellar mass of galaxies, $M_*$, but 
	 there is no correlation with specific SFR (SSFR), 
	 $\mbox{SFR}/M_*${, and dust attenuation $\ltir/\lfuv$}.  } 
	 {{Together, these
     results show that the AKARI FIS All-Sky Survey gives a
     representative sample of SF galaxies in the Local Universe.
     This sample will be a comprehensive standard of various
     properties of SF galaxies to be compared with, e.g., high-$z$
     SF galaxies.}
	 }
     \titlerunning{Star formation and dust extinction of galaxies from
       AKARI-GALEX}
\maketitle
%
%________________________________________________________________

\section{Introduction}

Star formation history of galaxies is one of the most important and
exciting topics in extragalactic astrophysics and observational
cosmology.  Especially, exploring the ``true'' absolute value of the
cosmic star formation rate (hereafter SFR) has been of a central
importance.

However, it remained difficult for a long time because of dust
extinction.  Even in the Local Universe, there aresome problems to 
estimate the absolute value of SFR
density because of different dependence of various SFR estimators on
dust attenuation \citep[e.g.,][and references therein]{hopkins06}.
Active star formation (SF) always comes along with dust production, 
because of various dust
grain formation processes related to the final stage of stellar
evolution \citep[e.g.,][]{dwek80,dwek98,nozawa03,takeuchi05c}.
Observationally, SFR of galaxies is measured by the ultraviolet (UV)
luminosity from massive stars because of their short lifetime ($\sim
10^8\;\mbox{yr}$) compared with the age of galaxies or the Universe.
However, the UV photons are easily scattered and absorbed by dust
grains.  
Hence the SFR of galaxies is always inevitably affected by dust which 
is produced by their SF activity.  
Since the absorbed energy is re-emitted
{at far-infrared (FIR) wavelengths, it is essential to
observe galaxies both at UV and FIR to have an unbiased view of their
SF \citep[e.g.,][]{buat05,seibert05,cortese06,takeuchi05a}.

{}To know the history of SFR in the Universe, we must know the
starting reference value, i.e., the SFR density in the Local Universe:
otherwise, we cannot define how much larger the SFR was in the past.  
However, since the volume of the Local Universe is limited by definition, an
all-sky survey is the only viable way to refine our knowledge of local
galaxies.
On the ``directly visible SF'' side, the advent of the UV surveyor-type 
satellite GALEX \citep{martin05} has changed the
situation of UV astronomy drastically.  
GALEX is performing an all-sky survey (All Sky Imaging Survey: AIS) 
at FUV (1530~\AA) and NUV (2300~\AA) with detection limits of 19.9 
and 20.8~mag \citep{morrissey07}, as well as deep surveys in some selected 
regions.
In a previous study, we have shown that GALEX AIS provides us with an 
unprecedented opportunity to explore the
visible SF in the Local Universe \citep{buat07a}.

As for the ``hidden'' side of SF, the Infrared Astronomical Satellite
\citep[IRAS; ][]{neugebauer84} has brought a vast amount of statistics
of dusty galaxies in the Local Universe by IRAS Point Source Catalog
(PSC) \citep[see, e.g.,][]{soifer87}.  Subsequently, FIR facilities
with much higher sensitivity have been launched, like ISO
\citep[e.g.,][]{genzel00,verma05} and Spitzer
\citep[e.g.,][]{soifer08}, and revealed the deep IR universe, but the
latter two were observatories dedicated to pointed observations.

In contrast to the latter two facilities, the Japanese IR satellite AKARI has 
performed various large-area surveys
at IR wavelengths \citep{murakami07} after IRAS, {\it especially including 
all-sky surveys at FIR and MIR}.  
In particular, with the
aid of the Far-Infrared Surveyor \cite[FIS:][]{kawada07} onboard,
various IR surveys were performed by AKARI.  
AKARI FIS has four FIR wavebands centered on $65, 90, 140$, and 
$160\;\mu$m, and a FIR all-sky survey was completed by this instrument.  
Since the latter two
bands are longer than $100\;\mu$m which was the longest wavelength
band of IRAS, {the obtained sample of dusty galaxies is less
biased than the IRAS sample, i.e., thanks to the better sensitivity to cooler dust than
IRAS, AKARI can detect galaxies with dust emission with lower temperatures.  
Thus, AKARI is a very promising facility to bring new knowledge of galaxies 
with cold dust.} 
In addition to these SF-related wavelengths, we need other various
wavelength bands to examine physical properties of galaxies, e.g.,
stellar mass [closely related to near-IR (NIR)], and intermediate
stellar population (related to optical).  For the former, we have a
set of all-sky data from 2-Micron All-Sky Survey
\citep[2MASS:][]{skrutskie06}, and for the latter, the SDSS final data
(DR7) are publicly available\footnote{URL: {\tt http://www.sdss.org/dr7/}.}, 
even if SDSS is not an all-sky survey.

In this work, we constructed a multiband galaxy catalog based on AKARI
All-Sky Survey 90-$\mu$m selected sources associated with IRAS PSC$z$
galaxies \citep{saunders00}.  Then, we measured GALEX FUV and NUV flux
densities, and associated SDSS and 2MASS photometries.
For this initial study, we have only selected ``secure'' galaxies with good 
photometric measurements for most of the bands.  
We present the sample construction in Section~\ref{sec:sample}.  
We describe basic properties of galaxies in
the sample in Section~\ref{sec:basic}.  Results on the SF and dust
attenuation are presented in Section~\ref{sec:results}.
Section~\ref{sec:conclusion} is devoted to our summary and
conclusions.

Throughout the paper we will assume $\Omega_{\rm M0} = 0.3$,
$\Omega_{\Lambda0} = 0.7$ and $ H_0 = 70 {\rm~ km~ s^{-1}~ Mpc^{-1}}$.
The luminosities are defined as $\nu L_{\nu}$ and expressed in solar
units assuming $L_{\odot} = 3.83 \times 10^{33} {\rm ~erg~ s^{-1}}$.

\begin{figure}[t]
\centering\includegraphics[width=0.4\textwidth]{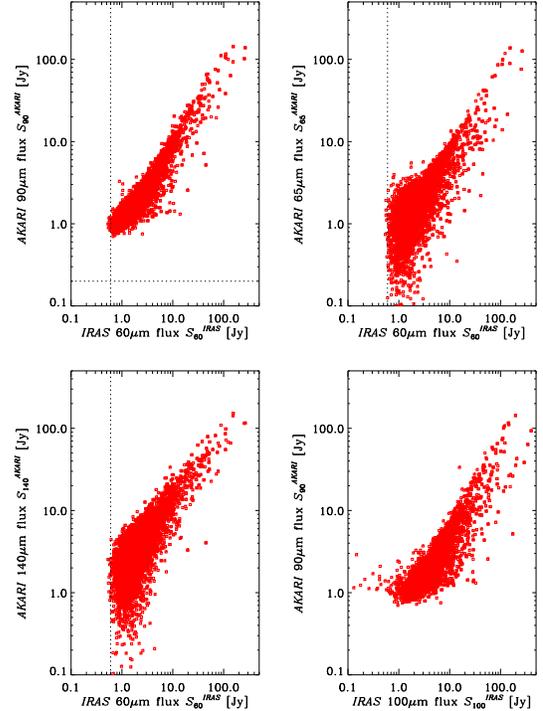}
\caption{Comparison between AKARI FIS and IRAS PSC$z$ flux
    densities.  
	{ Upper-left, upper-right, and lower-left
    panels present comparisons of IRAS $60\;\mu$m with AKARI
    $90\;\mu$m, $65\;\mu$m, and $140\;\mu$m flux densities of the
    AKARI-IRAS correlated sample.  
	The {vertical} dotted lines in
    these panels represent the flux density limit of IRAS PSC$z$.
    Lower-right panel shows a comparison of IRAS $100\;\mu$m 
	with AKARI $90\;\mu$m flux densities.
	{
	The horizontal dotted line in Upper-left panel represents the 
	formally expected detection limit of FIS 90~$\mu$m.
	}}
	}\label{fig:flux}
\end{figure}

\section{Sample construction}\label{sec:sample}

\subsection{Construction of the parent FIR sample from AKARI and IRAS PSC$z$}

\subsubsection{AKARI FIS All-Sky Survey}

The primary purpose of the AKARI mission is to provide
second-generation infrared (IR) catalogs with better spatial
resolution and wider spectral coverage than the IRAS catalog.
AKARI is equipped with a cryogenically cooled telescope of 68.5~cm
aperture diameter and two scientific instruments, the Far-Infrared
Surveyor \citep[FIS; ][]{kawada07} and the Infrared Camera \citep[IRC;
][]{onaka07}.  Among various astronomical observations performed by
AKARI, as we have mentioned in Introduction, an all sky survey with
FIS has been carried out (AKARI All-Sky Survey).  Since FIS is an
instrument dedicated to FIR $\lambda = 50 \mbox{--} 180\;\mu$m, all
the AKARI FIS bands are in the FIR wavelengths: {\it N60}
($65\;\mu$m), {\it WIDE-S} ($90\;\mu$m), {\it WIDE-L} ($140\;\mu$m),
and {\it N160} ($160\; \mu$m) \citep{kawada07}.  Hereafter, we
note as $S_{65}$, $S_{90}$, $S_{140}$ and $S_{160}$ the flux densities in
these bands.  
Especially, since FIS has sensitivity at longer
wavelengths than IRAS, a new classification scheme is needed
if we try to select a certain class of objects.  
Such a
scheme is not merely an empirical technique but also will provide us
with a new understanding of objects with cool dust which were
difficult to detect by IRAS bands.

We use the AKARI FIS Bright Source Catalogue (BSC), the first primary
catalog from the AKARI All-Sky Survey.  
Data from the $\beta$-1 version of this catalog are used in this work.  
AKARI BSC is supposed to have a uniform detection limit,
corresponding to per scan sensitivity, over the entire sky, except for
very bright sky parts where different data acquisition mode had to be
applied.  A summary of the All-Sky Survey is presented in \citet{yamamura09}.  
The AKARI FIS BSC provides data for 64311 sources.  
For each detected source,
AKARI source identifier, equatorial coordinates of the source position
and flux densities in the four FIR bands are given.  
Errors are not estimated for each individual source, but instead they are 
in total estimated to be 35~\%, 30~\%, 60~\%, and 60~\% at {\it N60}, 
{\it WIDE-S}, {\it WIDE-L}, and {\it N160}, respectively
\citep{yamamura08}.  
AKARI IRC performed another all sky survey, but the data are not fully reduced 
at the time we have been preparing this paper.  
Hence we focus on the FIS data only.

\subsubsection{Matching with IRAS PSC$z$}

The AKARI BSC contains many Galactic sources, like AGBs, H{\sc ii}
regions, planetary nebulae, etc. \citep[e.g.,][]{pollo10}. 
In order to construct a reliable
catalog of galaxies, we should pick up FIS sources confirmed as galaxies.  
For this purpose, we have performed a cross identification
of BSC sources with the IRAS PSC$z$ \citep{saunders00}.  
The PSC$z$ is a redshift survey of galaxies selected at IRAS $60\;\mu$m 
with a flux density limit of $S_{60} > 0.6\;\mbox{[Jy]}$\footnote{Because of this
step, we should note that we do not make a maximal use of the advantage of 
the long wavelength bands of AKARI FIS, since the sample is limited by IRAS 
bands ($\lambda < 100\;\mu$m).}.  
The PSC$z$ contains $\sim 16000$ galaxies.  
We have put a limit on recession velocity $v > 1000\;\mbox{km}\,\mbox{s}^{-1}$ 
so that we can avoid the effect of the peculiar velocity of galaxies.  
Then, we have matched the AKARI BSC
sources with PSC$z$ galaxies with a search radius of 36~arcsec, which
corresponds to the position uncertainty of IRAS PSC.  
The number of matched sources was 5890.  
{}To examine the effect of the choice of search radius, we changed the 
criterion from 20--60~arcsec.
This change of radius does not have a significant
impact on the resulting catalog ($\la 5$~\%).  
Hence, we use the catalog with a search radius of 36~arcsec in the following 
analysis.

For further analysis, we make a cross identification with SDSS
galaxies (see Sec.~\ref{subsection:sdss_2mass}).  
Then, we have to restrict our data only to the area covered by SDSS DR7
which is $8378.015\;\mbox{deg}^2$.
This restriction to this region of the sky has one advantage: the Galactic 
diffuse FIR emission is strong in some areas of the sky.  
In such regions, measured FIR flux
densities of point sources are contaminated by the Galactic emission
and not very accurate.  Since the SDSS region is selected so that the
Galactic extinction is small, the selected area automatically avoids
such FIR-bright regions.  Then, by this selection, our sample
automatically excludes strongly contaminated sources.  The resulting
parent catalog contains 1186 galaxies.

\subsubsection{Flux density comparison between AKARI and IRAS}

We compared AKARI and IRAS flux densities to examine our sample
selection.  The correlation is shown in Figure~\ref{fig:flux}.  The
horizontal dotted lines in upper-left, upper-right, and lower-left
panels represent the flux density limit of IRAS PSC$z$.  AKARI BSC
sources are selected at {\it WIDE-S}, i.e., $90\;\mu$m.  It is
important to see which selection controls the sample selection.  As
seen in the upper-left panel, the IRAS PSC$z$ limit and AKARI limit
are both well-defined, and neither of them strongly restricts the
sample.  The effective $90\;\mu$m flux density limit of our parent
sample is $\sim 0.8$~Jy.

\subsection{GALEX photometry}

GALEX AIS now observed $25000\;\mbox{deg}^2$ at FUV and NUV.  The
latest version of the public imaging is GR4/GR5\footnote{URL: {\tt http://galex.stsci.edu/GR4/}}.
We have measured the FUV and NUV photometry
of the parent AKARI galaxies as follows:
\begin{enumerate}
\item Cut out a $30' \times 30'$ square subimage from GALEX AIS images
  around each AKARI galaxy.
\item Select a subimage with the largest exposure time when multiple
  observations were available.
\item Measure FUV and NUV flux densities.  The NUV observation is
  taken as our reference.
\end{enumerate}
Since the sky coverage of GALEX AIS is not complete,
in some cases we do not have a proper
GALEX image for an AKARI galaxy.  In such a case we omit the
galaxy because we do not have any UV information.

Almost all of the sources are resolved by GALEX. 
They are thus very often separated into small bright patchy
regions, and the GALEX pipeline misidentifies these fragments as
individual objects.  
This is referred to as {\it shredding}.  We must
deal with the shredding to obtain sensible flux density measurements
for nearby extended galaxies.  
For this purpose, we have used an IDL
software package developed by ourselves.  
This software performs
aperture photometry in the NUV sub-image using a set of elliptical
apertures.  Total flux density is calculated within the aperture
corresponding to the convergence of the growth curve.  The sky
background is measured by combining several individual regions around
the source.  
NUV and FUV flux densities are corrected for Galactic
extinction using the Schlegel map \citep{schlegel98} and the Galactic
extinction curve of \citet{cardelli89}.  
A detailed description of the
photometry process can be found in \citet{iglesias06}.
 
This software was already used for our previous
IRAS-GALEX based studies \citep{buat07a}, and its performance is
carefully checked and established.
During} the procedure, we also excluded all the sources
contaminated by stars or too close to be disentangled by the
photometry software.  By this procedure, the number of remaining
galaxies is 776.

\subsection{Matching with SDSS and 2MASS galaxies}\label{subsection:sdss_2mass}

We further matched the AKARI-IRAS PSC$z$-GALEX sample (776 galaxies,
hereafter abbreviated as the AKARI-GALEX sample) with SDSS and 2MASS.

The AKARI-GALEX sample was matched with the 2MASS All-Sky Extended
Source (XSC) catalog in order to obtain the NIR ({\it J}, {\it H}, and
{\it Ks}) flux densities.  A matching radius of $20''$ was initially
adopted for the cross correlation.  All but 7 out of the 776 galaxies
in the AKARI-GALEX sample showed 2MASS counterparts at distances
closer than the matching radius.  In case of multiple candidates, we
always selected the brightest one at {\it Ks} band as the most
plausible one.  A further cross correlation with a matching radius of
$30''$ was attempted for these seven galaxies with no 2MASS
counterpart within the $20''$ radius.  However, again, no 2MASS
counterparts were found, and we adopted the standard 2MASS XSC upper
limits for these galaxies at {\it J}, {\it H}, and {\it Ks} bands.
The standard 2MASS limiting magnitudes at {\it J}, {\it H}, and {\it
  Ks} are 14.7, 13.9 and 13.1~mag, respectively \citep{jarrett00}.

We also matched the AKARI-GALEX sample to SDSS using the GALEX
coordinates for the AKARI-GALEX objects, and a search radius of
$15''$.  In the case of SDSS, large resolved galaxies such as those we
are dealing with here are shredded in multiple detections during the
deblending step of the pipeline.  We used the closest SDSS match to
the AKARI object to obtain the SDSS photometry of the parent object,
namely the object detected by the SDSS pipeline before deblending.  We
inspected all sources to check that the actual flux measured for the
parent object is not contaminated by nearby bright stars, artifacts
etc.

For most of the sample galaxies, SDSS galaxies were associated.
However, stars superposed on the SDSS galaxy images often hamper
accurate photometry.  It requires a very careful masking of the SDSS
image, and in the current analysis we simply omitted such galaxies.
After this selection, 607 galaxies remained.  We call this sample
``the final sample''.  Almost all galaxies have very secure
photometric data at UV, optical, NIR, MIR, and FIR, as well as the
redshift information.

\section{Basic Properties of the Sample Galaxies}\label{sec:basic}

\subsection{Number counts}\label{subsec:nc}

\begin{figure}[t]
%\centering\includegraphics[width=0.4\textwidth]{fig3a.ps}
\centering\includegraphics[width=0.4\textwidth]{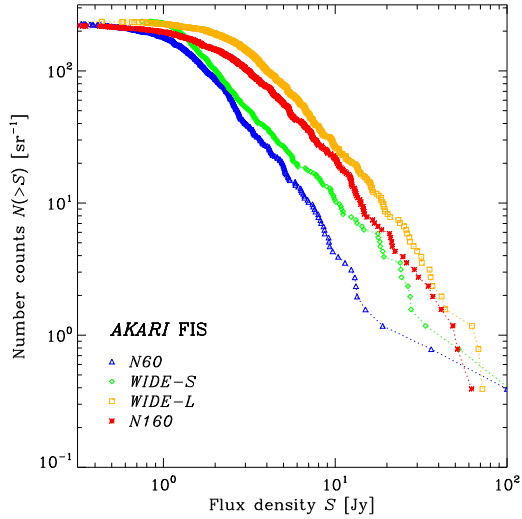}
\caption{Number counts of our AKARI multiband sample 
at AKARI FIS four bands.  
Triangles, diamonds, squares, and crosses represent AKARI
{\it N60}, {\it WIDE-S}, {\it WIDE-L}, and {\it N160} galaxy
counts, respectively.
}\label{fig:nc}
\end{figure}

Figure~\ref{fig:nc} presents the cumulative AKARI flux
density distribution (number counts) of our final sample.  We see that
the flux density limit of the sample at $90\;\mu$m is $\sim 1$~Jy,
shallower than the original limit of $\sim 0.8$~Jy, because
of additional constraints to have optical--NIR counterparts.

%IRAS $100\;\mu$m flux density distribution is similar to that of AKARI
%{\it WIDE-L} at their faint-end, but different at brighter regime.
%This may be because of the effect of confusion caused by the poorer
%angular resolution of IRAS.  
%Source confusion is important at the faintest flux densities around the detection limit of the survey
%\citep[][]{takeuchi04}, which does not seem to affect the sample flux
%measurements.  
%Since the discrepancy is larger at brighter flux
%regime, the source of confusion is very possibly Galactic cirrus, even
%we have tried to exclude such galaxies.  
%This is also consistent with
%the fact that the difference is larger at longer wavelengths.

\begin{figure}[t]
\centering\includegraphics[width=0.4\textwidth]{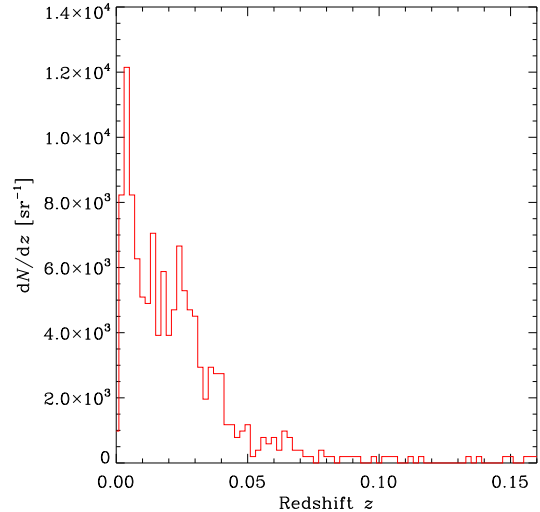}
\caption{Redshift distribution of the sample.
The distribution is normalized so that we obtain 
the number of galaxies per unit solid angle if we integrate 
it over redshifts.
}\label{fig:zdist}
\end{figure}

\begin{figure}[t]
\centering\includegraphics[width=0.4\textwidth]{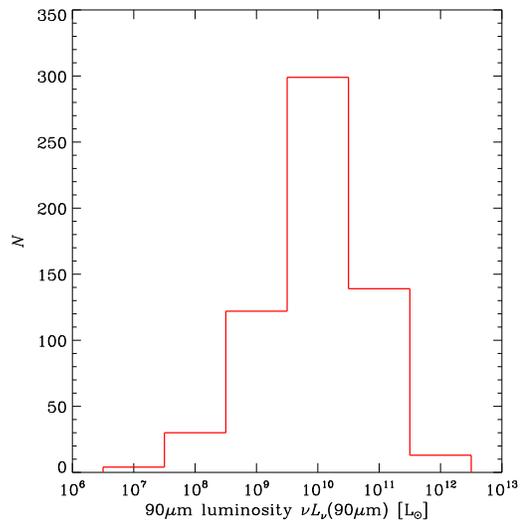}
\caption{Luminosity distribution of the sample at $90\;\mu$m.
}\label{fig:lumdist}
\end{figure}

\begin{figure}[t]
\centering\includegraphics[width=0.4\textwidth]{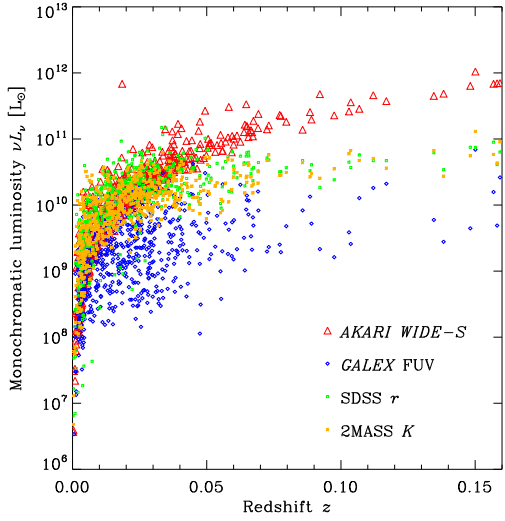}
\caption{Luminosity distribution of the sample at several wavelengths.
Triangles, diamonds, squares, and crosses represent galaxies AKARI 
{\it WIDE-S}, GALEX FUV, SDSS $r$, and 2MASS $K$-band, respectively.
}\label{fig:lum_compare}
\end{figure}

\subsection{Redshift and luminosity distributions}\label{subsec:zdist}

The redshift distribution of our final sample is
shown in Figure~\ref{fig:zdist}.  Since AKARI FIS BSC is rather
shallow, most of the sample locate at low redshifts $z \la 0.05$.
This is quite consistent with predictions of various IR galaxy
evolution models \citep[e.g.,][]{takeuchi01a,takeuchi01b,chary01}.

Figure~\ref{fig:lumdist} presents the 90-$\mu$m luminosity
distribution of the sample.  
Since this is a raw luminosity
distribution of IR galaxies, it decreases toward the faint end.  The
90-$\mu$m luminosity ranges from $10^6\;L_\odot \mbox{--}
10^{12}\;L_\odot$.  
Only a few galaxies are classified as ultraluminous IR galaxies (ULIRGs).  
The peak of the luminosity
distribution is around $10^{10} \; L_\odot$, which is more luminous
than the knee of the 60-$\mu$m luminosity function
\citep{takeuchi03a}.  
Since our selection procedure is multifold and
complicated, estimating a reliable luminosity function is not
straightforward.  
We will try this task in future works.

We can compare the distribution of galaxies on the
redshift--luminosity plane to see the effect of the shape of
luminosity function.  We show the $z$-$\nu L_\nu$ relation in
Figure~\ref{fig:lum_compare}.  Some representative wavelengths are
shown: AKARI {\it WIDE-S}, GALEX FUV, SDSS {\it r}, and 2MASS {\it
  Ks}.  We see that the luminosity at $90\;\mu$m increases
monotonically toward higher redshifts, while the optical and near-IR
(NIR) luminosities of the same galaxies saturate at certain values.
This is related to the difference in the shape of their luminosity
functions \citep[see, e.g.,][]{takeuchi05a,iglesias06}; i.e., the
luminous end of the function exponentially declines at optical-NIR and
UV, while it shows a power-law decline at FIR.  This is also related
to the fact that the more luminous galaxies are more strongly
extinguished by dust.  We will revisit this issue in a much more
direct way in Section~\ref{subsec:attenuation}

\begin{figure}[t]
\centering\includegraphics[width=0.4\textwidth]{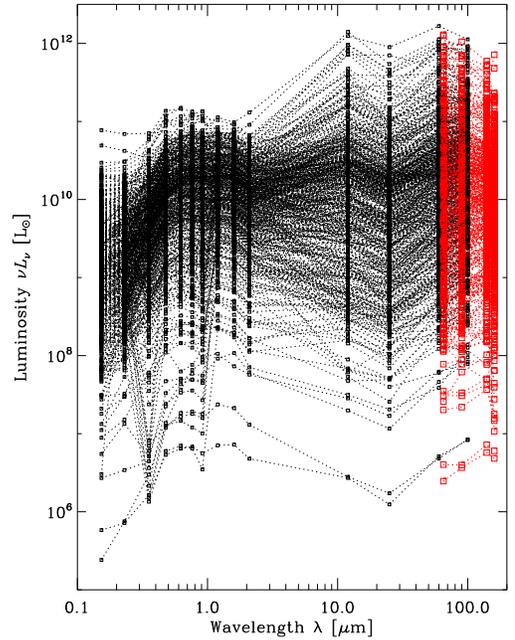}
\caption{Spectral energy distributions (SEDs) of the whole sample.
	Large empty squares represent the AKARI measurements, while the small 
	squares are data taken from GALEX FUV, GALEX NUV, SDSS {\it u}, {\it g}, 
	{\it r}, {\it i}, {\it z}, 2MASS {\it J}, {\it H}, {\it Ks}, and IRAS 12, 25, 60, and 
	$100\;\mu$m from left to right.
	{
	Dotted lines connect data points of each object to guide the eye.
	}
}\label{fig:sed}
\end{figure}

\begin{figure}[t]
\centering\includegraphics[width=0.4\textwidth]{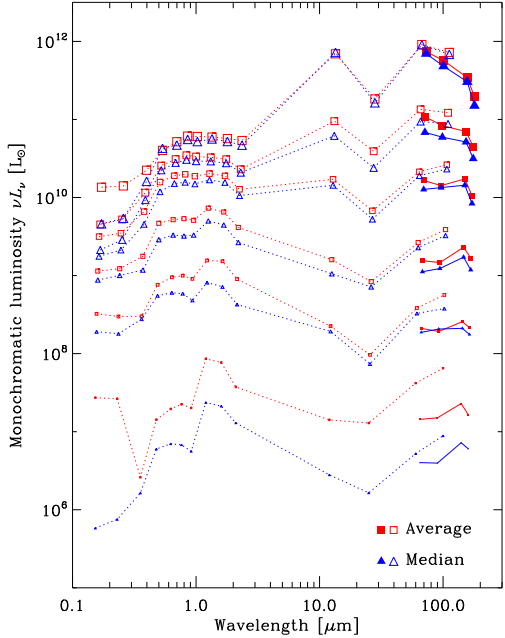}
\caption{Averaged spectral energy distributions of our sample as a
	function of $90\;\mu$m luminosity.
	Luminosity bins used
    here are the same as the bins in Figure~\ref{fig:lumdist}.  The
    central luminosities of the bins are $10^7$, $10^8$, $10^9$,
    $10^{10}$, $10^{11}$, and $10^{12}\;L_\odot$, respectively, and
    the obtained average SEDs are represented with different size of
    the symbols: from the smallest corresponding to
    $10^7 \;L_\odot$ to the largest corresponding
    to $10^{12}\;L_\odot$.  
	Filled and empty squares are averaged
	SEDs, while filled and empty triangles are median SEDs.
}\label{fig:avesed}
\end{figure}

\begin{figure}[t]
\centering\includegraphics[width=0.4\textwidth]{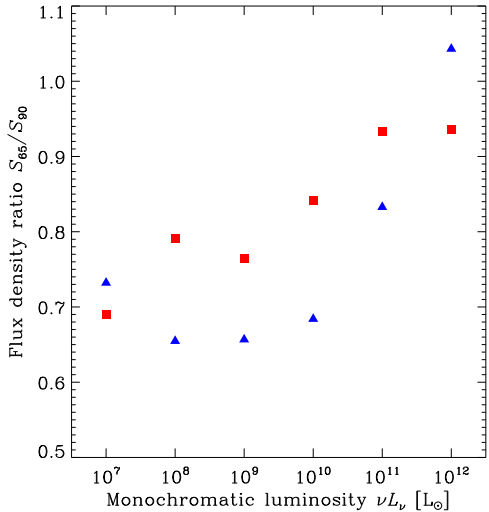}
\caption{
	Dependence of the flux density ratio
    $S_{65}/S_{90}$ on average 90-$\mu$m luminosity.
	Same as Fig.~\ref{fig:avesed}, the squares are
    estimated from averaged SEDs, while triangles are from median
    SEDs.
	}\label{fig:tsed}
\end{figure}

\subsection{SEDs of the sample}\label{subsec:sed}

Since we have the monochromatic luminosities from FUV to FIR, we
construct the spectral energy distributions (SEDs) of our sample.  We
show all the SEDs of the sample in Figure~\ref{fig:sed}.  Large empty
squares represent the AKARI FIS measurements, while the small squares
represent all the other data in Figure~\ref{fig:sed}.  We observe a
very large variety of SEDs among the sample galaxies.

{}To see a global trend of the SEDs more clearly, we sorted the
sample by their 90-$\mu$m luminosities and
subdivided the sample into six logarithmic bins from $10^6\;L_\odot$
to $10^{12}\;L_\odot$ (with a bin width $\Delta \log L_{90} = 1$).  
By taking an average and median, we constructed
``average SEDs'' as a function of 90-$\mu$m luminosity, shown in
Figure~\ref{fig:avesed}.  Because of a large dispersion and asymmetric
distribution of the SEDs in each bin, sometimes the average and median
SEDs do not agree very well.  Even so, now the trend is more clearly
seen: low 90-$\mu$m luminosity galaxies have cooler dust emission and
bluer UV-optical continuum, while high 90-$\mu$m luminosity galaxies
have hotter dust emission and redder UV-optical continuum.

The 90-$\mu$m luminosity dependence of the dust emission
temperature is more clearly seen if we plot a flux density ratio
$S_{65}/S_{90}$ as a function of $L_{90}$.  This is shown in
Figure~\ref{fig:tsed}.  
Again, the agreement between average and
median values is not excellent, we see a
clear monotonically increasing trend of the flux density ratio along
with $L_{90}$.  

The lowest luminosity galaxies have the largest uncertainty mainly
because of poor statistics.  
This problem will be solved by constructing deeper and larger sample, 
possibly by the next generation AKARI FIS catalog.  
The most 90-$\mu$m luminous galaxies seem to have an upturn at UV.  
This may be because of an AGN component in these galaxies.  
We will examine this issue in our future work.  
In both Figures~\ref{fig:sed} and \ref{fig:avesed}, it is clear that AKARI
FIS measurements at {\it N60} and {\it WIDE-S} do not agree with
IRAS ones. 
This discrepancy is mostly due to the better angular
resolution of AKARI FIS compared with IRAS.  
Thus, the measured values are much less contaminated by
Galactic cirrus emission.  Also because of better angular
resolution, the source confusion effect is much smaller than
for IRAS.
As both cirrus and source confusion effects cause flux boosting for IR
galaxy number counts, the IRAS flux densities could be overestimated
\citep[][]{takeuchi04}.  
\citet{jeong07} also examined the same
problem with an early AKARI sample and concluded
that the difference between IRAS and AKARI flux densities are due to
the confusion effects.
We see a strange bump at $12\;\mu$m for luminous galaxies.  
This may not be a physical feature in the SEDs of
our sample but because of a poor measurement at this band by IRAS.  
In the redshift--luminosity diagram at $12\;\mu$m
(see Fig.~\ref{fig:lum}), we see that many galaxies locate on
the limiting luminosity line.  
This means that, even if they are
classified as measurements, actually they are upper limits of the flux
density.  We plan to study this point further by AKARI IRC all-sky
survey in the future.

\subsection{The $\mbox{NUV}-r$ distribution}

\begin{figure}[t]
\centering\includegraphics[width=0.4\textwidth]{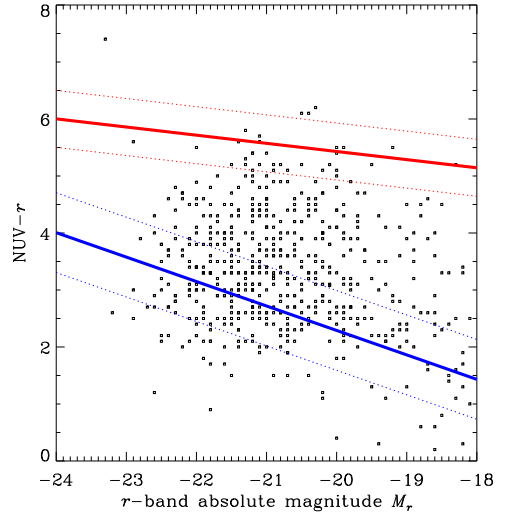}
\caption{The absolute $r$-magnitude-$\mbox{NUV}-r$ color distribution. 
{Upper and lower thick solid lines represent the average red sequence 
and blue cloud taken from \citet{salim07}, with its envelope indicated by 
dotted lines.
}
}\label{fig:nuvr}
\end{figure}

The $\mbox{NUV}-r$ restframe color is very efficient to separate the
galaxies into blue and red populations \citep{salim07,martin07}.  At
high-$z$, IR selected galaxies observed with Spitzer were found to
populate the so-called green valley between the blue cloud and red
sequence, first mentioned by \citet{bell05} and further identified in
the COSMOS fields \citep{kartaltepe09,vergani09}.

However, in contrast, no systematic study of IR selected galaxies has
been performed in the nearby universe until now.
Figure~\ref{fig:nuvr} shows the distribution of the $\mbox{NUV}-r$
color for our sample galaxies against the absolute magnitude at
$r$-band, $M_r$.  The distribution in $\mbox{NUV}-r$ is found to be
large and monolithic in contrast to the bimodal distribution found in
optical surveys.  
{ {}From Figure~1 of \citet{salim07} (based on an SDSS-GALEX 
selected sample), we approximate the red sequence on this diagram with 
a linear relation
\begin{eqnarray}
  \mbox{NUV} - r = -0.143M_r + 2.57 \;,
\end{eqnarray}
with $\pm 0.5$ envelopes, and the the blue cloud with
\begin{eqnarray}
  \mbox{NUV} - r = -0.429M_r - 6.29\;,
\end{eqnarray}
with $\pm 0.7$.  
Clearly, we see that our galaxies lie in the blue
cloud and populate the green valley to produce a continuous
distribution, and only a few objects are located in the red sequence.
} 
Therefore, the color distribution of FIR-selected galaxies is different than 
the one obtained from an optical selection, as they preferentially populate 
the green valley between the red sequence and blue cloud.  
An interpretation of this behavior in terms of dust attenuation, SF history 
or AGN activity will be developed in a future paper.

\section{Star Formation and Dust Extinction of the $90\;\mu$m-Selected Galaxies}\label{sec:results}

\begin{figure*}[t]
\centering\includegraphics[width=0.4\textwidth]{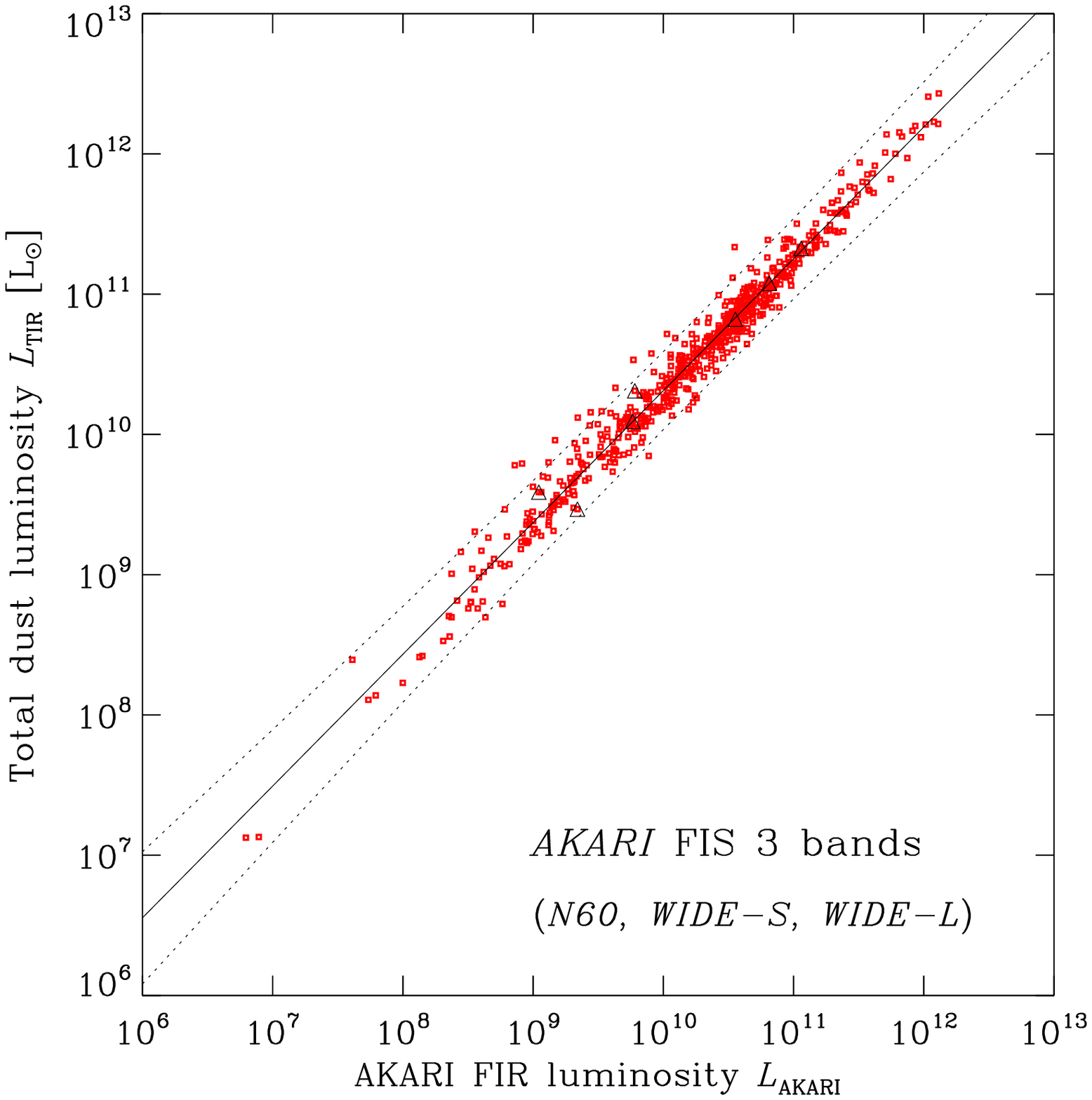}
\centering\includegraphics[width=0.4\textwidth]{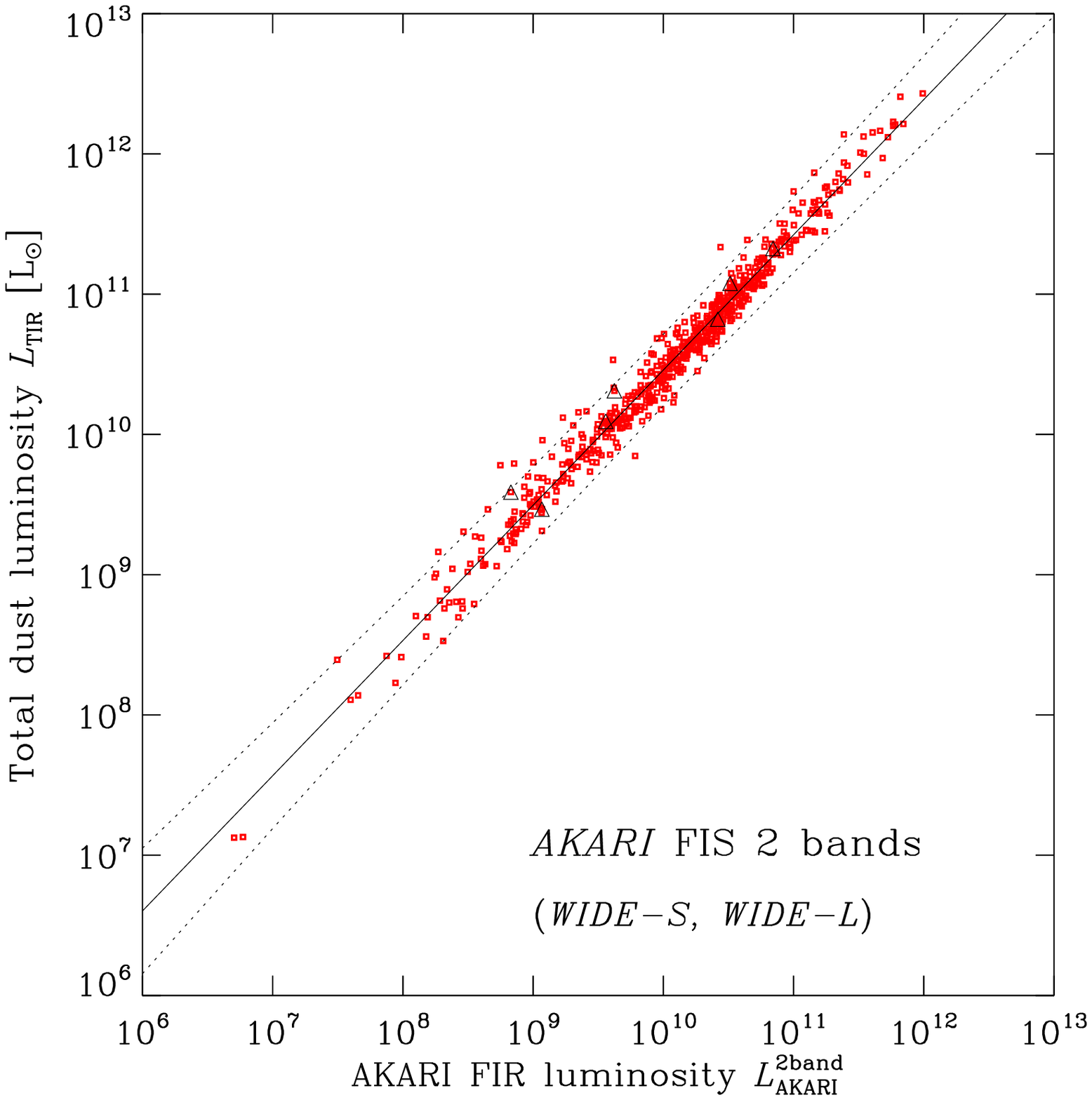}
\caption{Relation between the AKARI FIR luminosity and total IR (TIR) luminosity of the sample.
{Left panel is for the AKARI FIR luminosity defined by eq.~(\ref{eq:lum_akari}), while right panel
is for the AKARI FIR luminosity from 2 bands, defined by eq.~(\ref{eq:lum_akari2}). 
Black solid lines are the best least-square fits to the data, and the dotted curves represent
the 95~\% confidence levels of the lines.
Triangles are galaxies with IRAS $100\;\mu$m quality flag larger than 3, 
i.e., those with insecure flux density measurement.}
}\label{fig:fir}
\end{figure*}

\subsection{Total dust luminosity from AKARI FIS bands}\label{sec:ltir}

{}To calculate SF- and attenuation-related physical properties of
the galaxies, a total IR (TIR) luminosity is
required.  Various estimators of $\ltir$ were proposed by previous
authors \citep[][]{helou88,dale01,dale02, sanders96}.
\citet{takeuchi05b} has shown that we can safely estimate the TIR
luminosity by a formula proposed by \citet{sanders96}:
\begin{eqnarray}\label{eq:sanders}
  \ltir &\equiv& 4.93 \times 10^{-22} [ 13.48 L_\nu(12\mu\mbox{m})+5.16L_\nu(24\mu\mbox{m})
    \nonumber \\
    &&+2.58L_\nu(60\mu\mbox{m}) + L_\nu(100\mu\mbox{m})] \; [L_\odot].
\end{eqnarray}
Since our sample contains all IRAS band flux densities, we can use the
estimator of Sanders \& Mirabel to examine the AKARI FIR flux. 
In this subsection, we make an attempt to establish a reliable formula to 
convert a FIR luminosity measured by AKARI to the TIR luminosity.

Since AKARI FIS has continuous bands from $\sim 50\;\mu$m to $\sim
160\;\mu$m, we can easily define FIR flux simply by adding the flux
densities multiplied with their bandwidths (in [Hz]).
\citet{hirashita08} has defined the AKARI FIR luminosity as
\begin{eqnarray}\label{eq:lum_akari}
  \lakari &=& \Delta \nu \left( \mbox{\it N60} \right) L_\nu (65\;\mu\mbox{m})+
    \Delta \nu \left( \mbox{\it WIDE-S} \right) L_\nu (90\;\mu\mbox{m})\nonumber \\
    &&+\Delta \nu \left( \mbox{\it WIDE-L} \right) L_\nu (140\;\mu\mbox{m}) \;,
\end{eqnarray}
where
\begin{eqnarray}
  &&\Delta \nu \left( \mbox{\it N60} \right) = 1.58 \times 10^{12} \mbox{[Hz]}  \\
  &&\Delta \nu \left( \mbox{\it WIDE-S} \right) = 1.47 \times 10^{12} \mbox{[Hz]}  \\
  &&\Delta \nu \left( \mbox{\it WIDE-L} \right) = 0.831 \times 10^{12} \mbox{[Hz]}  \; .
\end{eqnarray}
However, since the sensitivity of AKARI {\it N60} is not as good as
other two wide bands, if we can use FIR luminosity defined only by
{\it WIDE-S} and {\it WIDE-L} , it will be useful because we can have
larger number of galaxies.  We define $\lakari^{\rm 2band}$ by
omitting the term of {\it N60} as
\begin{eqnarray}\label{eq:lum_akari2}
  \lakari^{\rm 2band} &=& \Delta \nu \left( \mbox{\it WIDE-S} \right) L_\nu (90\;\mu\mbox{m})+\nonumber \\
   && \Delta \nu \left( \mbox{\it WIDE-L} \right) L_\nu (140\;\mu\mbox{m}) \;.
\end{eqnarray}

Figure~\ref{fig:fir} presents correlations between $\lakari$ and
$\ltir$.  Left panel shows the correlation between $\lakari$ of
\citet{hirashita08} and $\ltir$, while right panel shows the one
between $\lakari^{\rm 2band}$ and $\ltir$.  The fitting results are as
follows:
\begin{eqnarray}\label{eq:fir1}
  \log \ltir &=& 0.940 \log \lakari + 0.914 \;, \\
  r &=& 0.987 \;,
\end{eqnarray}
and 
\begin{eqnarray}\label{eq:fir2}
  \log \ltir &=& 0.964 \log \lakari^{\rm 2band} + 0.814 \;, \\
  r &=& 0.989 \;,
\end{eqnarray}
where $r$ is the correlation coefficient.  The solid lines in
Figure~\ref{fig:fir} depict these equations.  The envelopes delineated
by dotted lines in Figure~\ref{fig:fir} represent the 95~\% confidence
intervals.

In some cases the quality of IRAS
$100\;\mu$m flux density measurement is poor; we examine the impact 
of these objects in Figure~\ref{fig:fir}.
Triangles represent galaxies with IRAS
$100\;\mu$m quality flag larger than 3, i.e., the ones for which the
measurement was difficult and not secure.  
However, even so, clearly
they lie well within the distribution of the whole sample and do not
have a significant impact on the statistical analysis.

Since these plots present luminosity-luminosity correlations, the very
tight correlations are not extremely surprising.  However,
uncertainties of these equations are within a factor of $\sim
2\mbox{--}3$, which means that the estimation works very well even if
we do not have MIR measurements.  We also stress that the correlation
is even better if we only use AKARI wide bands.

Hereafter, we use the total IR luminosity $\ltir$ obtained from AKARI
{\it N60}, {\it WIDE-S}, and {\it WIDE-L} [equation~(\ref{eq:fir1})]
for all the following analysis.  {\it Namely, when we mention $\ltir$,
  it is always AKARI-based total IR luminosity.}

\begin{figure}[t]
\centering\includegraphics[width=0.4\textwidth]{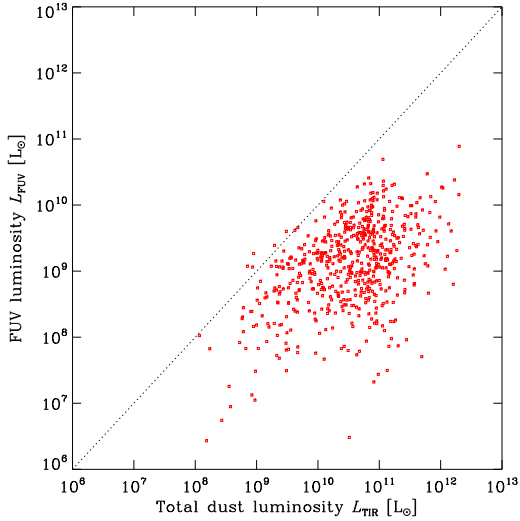}
\caption{Comparison between TIR and FUV luminosities. 
{The diagonal dotted line represents the case if $L_{\rm TIR}$ equals $L_{\rm FUV}$.
}
}\label{fig:uv_fir}
\end{figure}

\subsection{Luminosity from current star formation}\label{subsec:lsf}

Now we can compare the FUV ($\lfuv$) and
total IR (TIR) ($\ltir$) luminosities for our final sample.  
This is shown in Figure~\ref{fig:uv_fir}.  
It is striking that the luminosity is dominated by $\ltir$ for the vast majority 
of our sample, even though we take into account that they are FIR-selected.
Also, it is worth mentioning that the luminosity at FUV does not
exceed $10^{11}\;L_\odot$.  In contrast, $\ltir$ can be much higher.

\begin{figure*}[t]
\centering\includegraphics[width=0.4\textwidth]{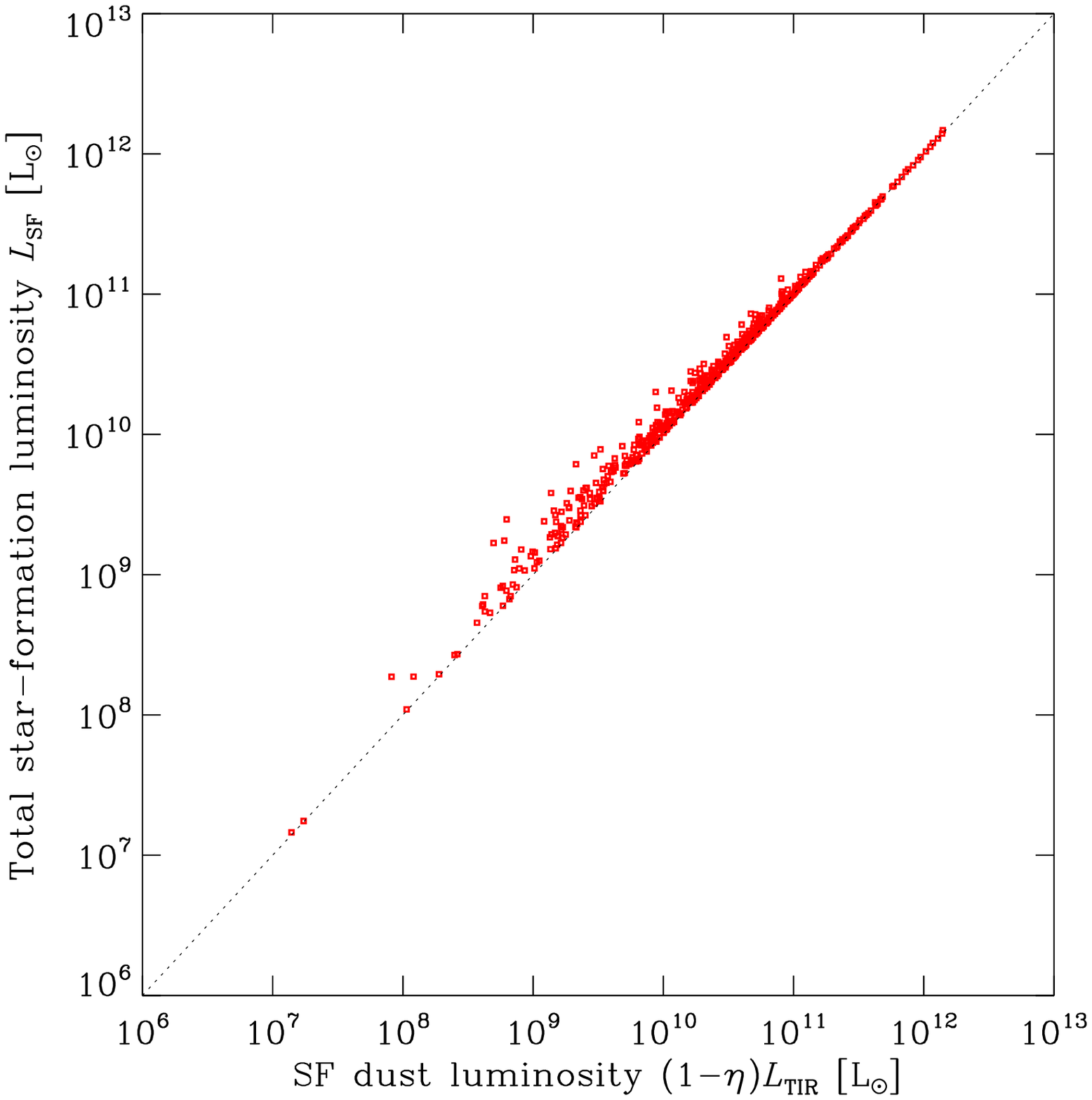}
\centering\includegraphics[width=0.4\textwidth]{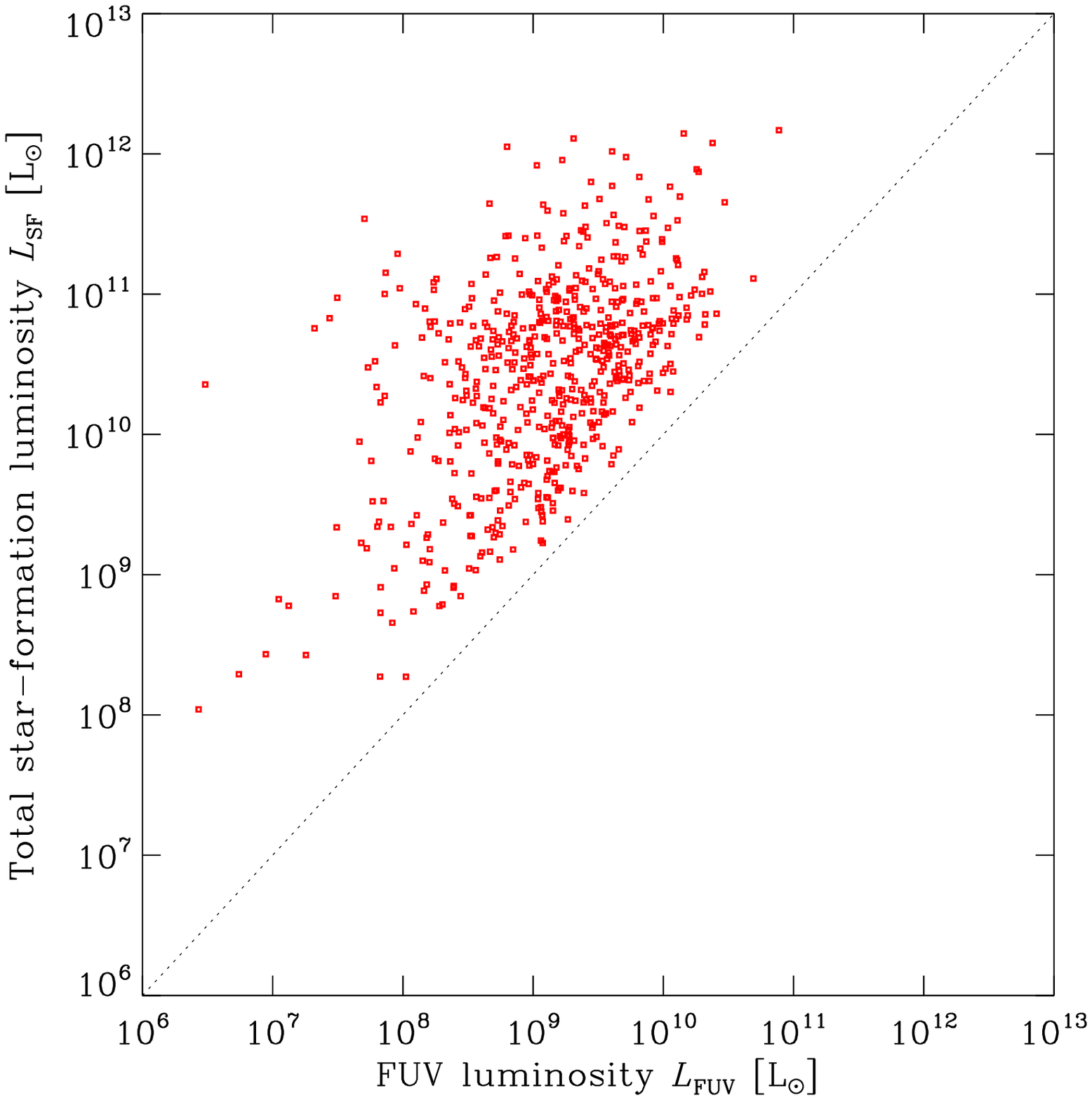}
\caption{
Contribution of $\ltir$ and $\lfuv$ to the total star formation (SF) luminosity $\lsf$,
{which is the luminosity produced by newly forming stars defined by eq.~(\ref{eq:lsf}).
In the left panel, $\eta$ is the fraction of IR emission produced by dust heated by 
old stars which is not related to the current SF (see main text).
}
}\label{fig:lsf}
\end{figure*}

By combining $\lfuv$ and $\ltir$, we can define the luminosity
contribution from newly formed stars, $\lsf$:
\begin{eqnarray}\label{eq:lsf}
  \lsf \equiv \lfuv + (1-\eta) \ltir \;,
\end{eqnarray}
where $\eta$ is the fraction of the dust emission due to the heating
of grains by old stars which is not directly related to the recent
SFR.  We adopt a value of 30~\% for this fraction \citep{hirashita03}.
The contribution of $\lfuv$ and $\ltir$ is shown in
Figure~\ref{fig:lsf}.  Naturally, the contribution of $(1-\eta)\ltir$
dominates $\lsf$.  In contrast, the contribution of $\lfuv$ has a very
large scatter, and the correlation is very poor.
Hence, it is not surprising that it is very difficult to estimate the total energy 
emitted by newly formed stars only from FUV information, even in an average
sense.  
The UV contribution becomes significant at lower
luminosities $\lsf < 10^{10}\;L_\odot$, as seen in the left panel of
Figure~\ref{fig:lsf}.

\subsection{Star formation rate of galaxies in our sample}\label{subsec:sfr}

Here, we interpret the data in terms of SFR.  Assuming a constant SFR
over $10^8$~yr, and Salpeter initial mass function (IMF) \citep[][mass
range: $0.1\mbox{--}100\;M_\odot$)]{salpeter55}, we have the relation
between the SFR and $\lfuv$
\begin{eqnarray}
  \log \sfr_{\rm FUV} = \log \lfuv  - 9.51 \;.
\end{eqnarray}
For the IR, to transform the dust emission to the SFR, we assume that
all the stellar light is absorbed by dust.  Then, we obtain the
following formula under the same assumption for both the SFR history
and the IMF as those of the FUV,
\begin{eqnarray}
  \log \sfr_{\rm dust} = \log \ltir - 9.75 + \log (1-\eta) \;.
\end{eqnarray}
Here, again, $\eta$ is the fraction of the dust emission by old stars.
Thus, the total SFR is simply
\begin{eqnarray}
  \sfr = \sfr_{\rm FUV} + \sfr_{\rm dust} \;.
\end{eqnarray}

\begin{figure}[t]
\centering\includegraphics[width=0.4\textwidth]{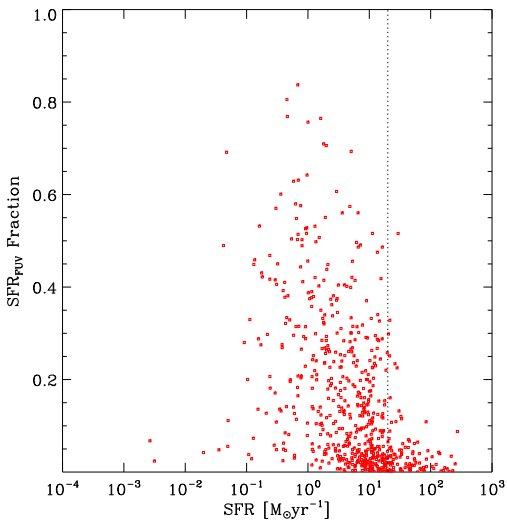}
\caption{
Contribution of the FUV-estimated (or ``directly visible'') SFR, $\sfr_{\rm UV}$, to 
the total SFR as a function of total SFR.
{Vertical dotted line represents the effective boundary above which almost
all energy produced by newly forming stars is emitted at IR.
}
}\label{fig:sfr}
\end{figure}

The obtained SFR is shown as a function of the fraction of the
contribution of $\sfr_{\rm FUV}$ in Figure~\ref{fig:sfr}.  Reflecting
the large scatter of $\lfuv/\lsf$, the scatter of $\sfr_{\rm
  FUV}/\sfr$ is very large at $\sfr <
20\;\mbox{M}_\odot\,\mbox{yr}^{-1}$.  However, quite sharply, no
galaxies have a large contribution of $\sfr_{\rm UV}$ at $\sfr >
20\;\mbox{M}_\odot\,\mbox{yr}^{-1}$.  
The vertical dotted line shows
this ``threshold'' SFR in Figure~\ref{fig:sfr}.

\subsection{Dust attenuation}\label{subsec:attenuation}

\subsubsection{Dust attenuation of the sample}

\begin{figure}[t]
\centering\includegraphics[width=0.4\textwidth]{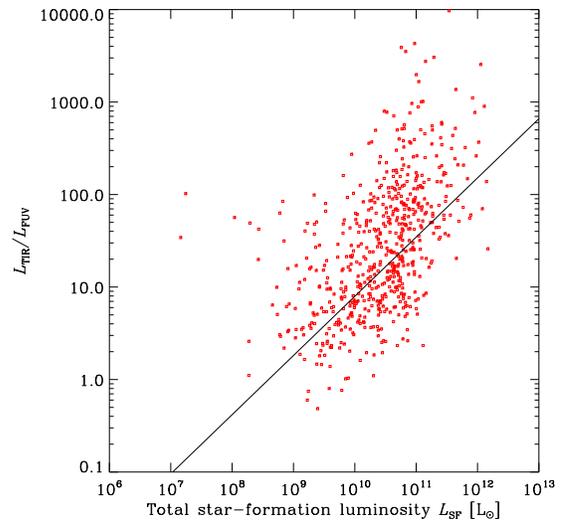}
\caption{Distribution of $\ltir$-$\lfuv$ ratio as a function of
  the star formation luminosity $\lsf$.
  {Solid line is the relation found by IRAS-GALEX analysis
  \citep{buat07a}, represented by eq.~(\ref{eq:localdust}).
}
}\label{fig:afuv}
\end{figure}

Galaxies selected in IR are expected to have a quite large dust
attenuation.  Here we examine the extinction properties of galaxies in
the sample.  For this study, a good observational indicator of dust
attenuation is required.  
The $\ltir/\lfuv$ ratio is widely recognized
to be a robust measure of dust attenuation.  This ratio was found to
increase with the star formation luminosity in a similar way from
$z=0$ to $z=0.7$ \citep{martin05,buat07b,zheng07}. 
Dust attenuation can be derived using the formula of \citet{buat05}:
\begin{eqnarray}
  A_{\rm FUV} &=& -0.0333~\left(\log \frac{L_{\rm TIR}}{L_{\rm FUV}}
   \right)^3 \nonumber \\
   &&+0.3522~\left(\log\frac{L_{\rm TIR}}{L_{\rm FUV}}\right)^2\nonumber \\
   &&+1.1960~\left(\log\frac{L_{\rm TIR}}{L_{\rm FUV}}\right)+0.4967 \; [\mbox{mag}]\;.
\end{eqnarray}
Figure~\ref{fig:afuv} presents the variation of $\ltir/\lfuv$ as a
function of the star formation luminosity.  A clear increase of
$\ltir/\lfuv$ with $\lsf$ is seen.  The solid line is the mean trend
of local galaxies found in our IRAS study \citep{buat07a}:
\begin{eqnarray}\label{eq:localdust}
  \log \left( \frac{\ltir}{\lfuv}\right) = 0.64 \log \lsf - 5.5 \;.
\end{eqnarray}
This line was estimated by a weighting of $\ltir/\lfuv$ with $1/V_{\rm
  max}$ to eliminate the flux selection effect, i.e., the effect that
more luminous galaxies can be more easily detected.  In contrast, we
plot raw values of $\ltir/\lfuv$.  However, though the steep rise of
the distribution of $\ltir/\lfuv$ with $\lsf$ is partially because of
this selection effect, the trend is still well consistent with our
IRAS study \citep{buat07a}, and the conclusion would remain valid.

\subsubsection{{Dust attenuation versus the slope of the UV continuum}}

The slope of the UV continuum is commonly used as a proxy to estimate
dust attenuation and $\ltir/\lfuv$ when IR data are not available or
considered as unreliable \citep{daddi07,reddy08,reddy09}.  This method
is based on a calibration performed on local starburst galaxies
\citep{meurer95,meurer99}.  
GALEX observations brought large amount of
data in this field since the slope of the UV continuum can be safely
deduced from the $\mbox{FUV}-\mbox{NUV}$ color.  Studies based on
GALEX data have shown that the local starburst calibration is not
valid for the bulk of local star forming galaxies
\citep[e.g.,][]{dale05,cortese06,boissier07,gildepaz07,johnson07}:
i.e., these galaxies exhibit lower dust attenuation ($\ltir/\lfuv$ )
than expected from their $\mbox{FUV}-\mbox{NUV}$ color.  IR selected
galaxies also depart from the starburst law but in the opposite way:
they are found to exhibit a $\ltir/\lfuv$ ratio much larger than
expected \citep{buat05,goldader02}.  
Here, we re-investigate this
issue making use of a much larger sample than used in \citet{buat05}.

\begin{figure*}[t]
\centering\includegraphics[width=0.4\textwidth]{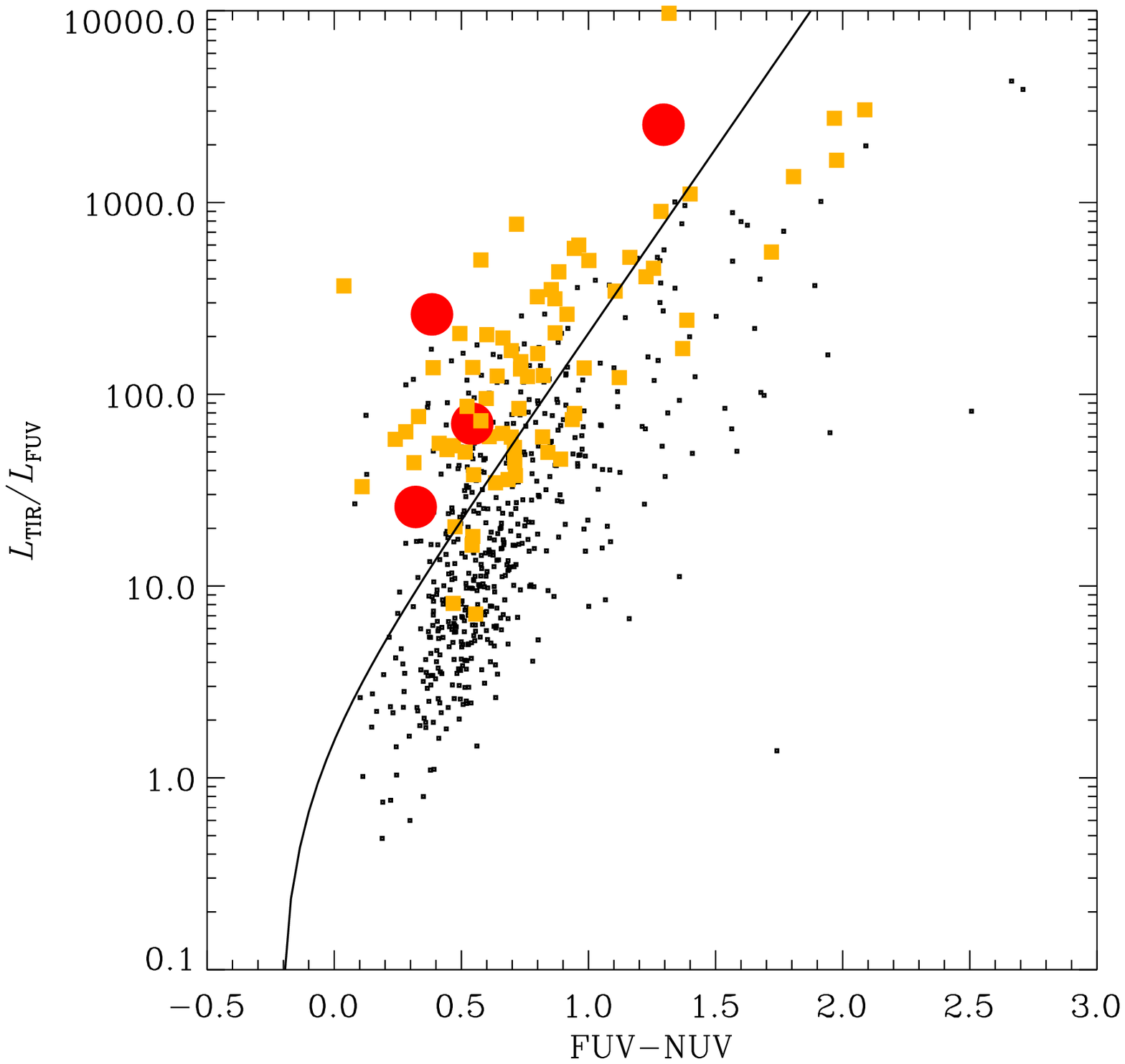}
\centering\includegraphics[width=0.4\textwidth]{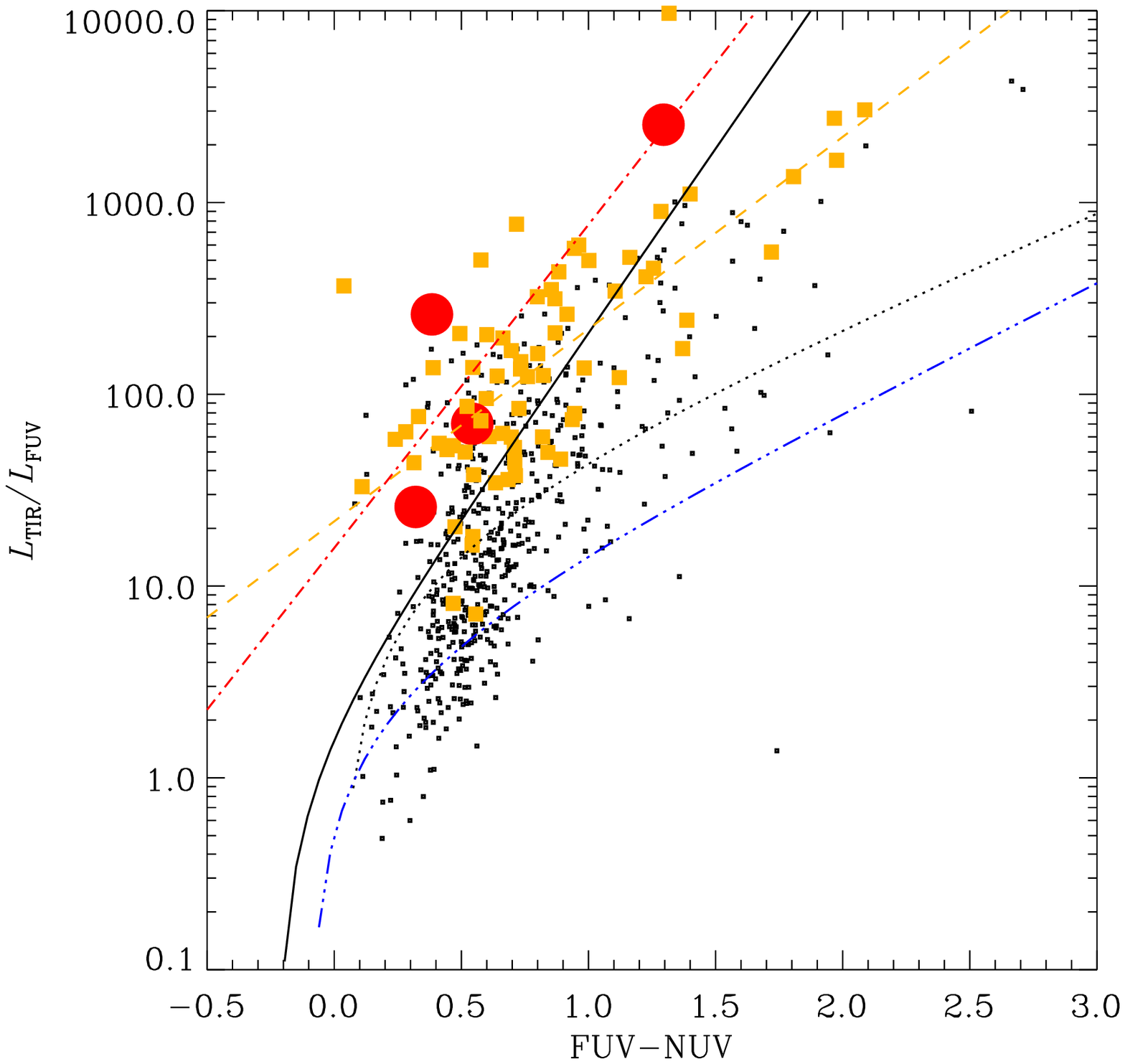}
\caption{
Distribution of $\ltir$-$\lfuv$ luminosity ratio as a function of UV color $\mbox{FUV}-\mbox{NUV}$.
Dots: galaxies with $60\mu$m luminosity $L_{60} < 10^{11}\;L_\odot$, 
filled squares: galaxies with $10^{11} \le L_{60} < 10^{12}\;L_\odot$ (IR luminous galaxies: LIRGs), 
and filled circles: galaxies with $10^{11} \le L_{60} < 10^{12}\;L_\odot$ 
(Ultraluminous IR luminous galaxies: ULIRGs).
{
Solid lines in each panel represent the relation for Local starbursts
proposed by \citet{meurer99}, which was converted into the relation 
between the FUV-NUV color [eq.~(\ref{eq:meurer})].
}
Left panel shows the raw distribution of $\ltir$-$\lfuv$ luminosity ratio, while
in Right panel functional fits are overplotted on the data. 
For galaxies with $L_{60} < 10^{11}\;L_\odot$, eq.~(\ref{eq:meurer}) is used, while
for more luminous galaxies, simple linear relations are adopted because of poor
fit of the function.
Dotted line: a fit to galaxies with $L_{60} < 10^{11}\;L_\odot$ by
eq.~(\ref{eq:kong});
Dot-dot-dot-dashed line: a fit to a sample of normal galaxies by eq.~(\ref{eq:kong})
presented by \citet{boissier07};
Dashed line: a linear fit to LIRGs;
Dot-dashed line: a linear fit to ULIRGs.
}\label{fig:irx}
\end{figure*}

Figure~\ref{fig:irx} shows the FUV-NUV color against
$\ltir/\lfuv$ (often called the IR-excess (IRX)--$\beta$ relation) for
our sample galaxies together with the local starburst relation.
{The local relation was taken from \citet{meurer99} and
  converted into the relation between the FUV-NUV color and
  $\ltir/\lfuv$ by eqs.~(1) and (2) of \citet{kong04}:
\begin{eqnarray}\label{eq:meurer}
  \frac{\ltir}{\lfuv} = 10^{\left[ 1.92({\rm FUV} - {\rm NUV}) + 0.4\right]} -0.95 \;.
\end{eqnarray}
} Most of the galaxies lie below the starburst relation by a factor of
5 to 10, but some locate above the relation.  A significant fraction
of galaxies above the local starburst line are luminous IR galaxies
(LIRGs and ULIRGs).  
Especially, all the ULIRGs have larger
$\ltir/\lfuv$ ratios than expected from the relation.
This general trend is quite consistent with \citet{buat05}.  
It may be
worth mentioning that LIRGs are consistent with the local starburst line.  
{These trends are more clearly represented by
functional fits in Right panel of Figure~\ref{fig:irx}.  
We used a functional form of eq.~(\ref{eq:meurer}) for ``normal'' galaxies 
with $L_{60} < 10^{11}\;L_\odot$, 
\begin{eqnarray}\label{eq:kong}
  \frac{\ltir}{\lfuv} = 10^{\left[ a({\rm FUV} - {\rm NUV}) + b\right]} -c \;.
\end{eqnarray}
with letting three parameters,  $a, b$ and $c$ free.
The fitted relation is well below the local starburst relation originally claimed 
by \citet{meurer99}.  
For more luminous galaxies, eq.~(\ref{eq:kong}) does not give a reasonable
fit, hence we adopted a simple linear fit.  
Even if LIRGs are
roughly consistent with the local starburst relation, the fitted
line is much flatter than that, because of the existence of galaxies
well above the local starburst line with blue UV color.  
It is difficult to conclude in the case of ULIRGs because of poor statistics, 
but the fitted linear relation seems to locate above the local 
starburst relation.
}
{
We have seen that galaxies with relatively quiescent galaxies ($L_{60} < 10^{11}
\;L_\odot$) in our sample locate well below the relation of \citet{meurer99}.
Similar trend was reported for optically selected galaxies 
\citep[e.g.,][]{boissier07,cortese06}
\citet{boissier07} investigated this relation for normal galaxies selected at 
optical wavelength with the same functional form with this study 
[eq.~(\ref{eq:kong})]
They found a much flatter relation than Meurer's original relation, which
is presented in Right panel of Figure~\ref{fig:irx} with dot-dot-dot-dashed
line.
\citet{cortese06} reported a similar result.
What is newly clarified in this study is that the same trend is seen for 
purely FIR-selected sample of SF galaxies.
In addition, Boissier et al.'s relation does not represent the relation for
our quiescent galaxy sample. 
This is naturally understood that our sample includes strongly reddened
galaxies by dust which can never be picked up by UV-selection.
}

\subsubsection{Attenuation and stellar mass-related quantities}

By using optical--NIR flux densities, we can estimate the stellar mass
of the sample galaxies.  Here we discuss dust attenuation properties
with respect to the stellar mass.  There are various methods to
estimate stellar mass.  In this work, we used an SDSS-based method
proposed by \citet{yang07}; these authors assume the Kroupa IMF
\citep{kroupa01}, while we used the Salpeter IMF \citep{salpeter55}
to estimate the SFR of galaxies.
For consistency, we convert their stellar mass estimates using the
conversion factor given by \citet{bell03}.  As for the accuracy of the
stellar mass estimates, the relation has been calibrated from
SDSS+2MASS data, and the error should be around 20~\%.  We also
estimated the stellar mass from $Ks$-band using the relations derived
by \citet{bell03}, and the trends did not change significantly.
Further exploration for the stellar mass estimation will be discussed 
elsewhere.

\begin{figure}[t]
\centering\includegraphics[width=0.4\textwidth]{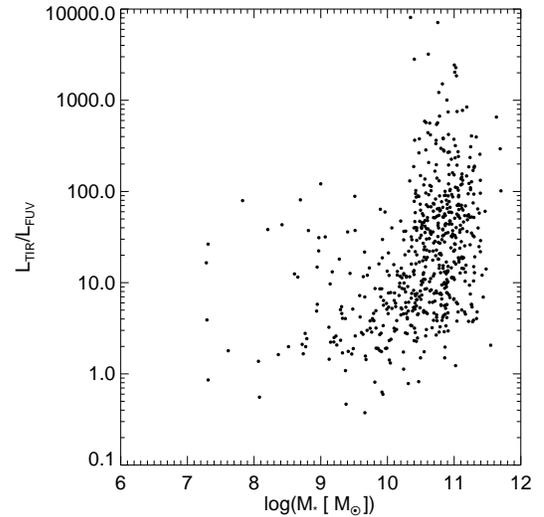}
\caption{
Relation between stellar mass $M_*$ and dust attenuation indicator $\ltir/\lfuv$.
}\label{fig:afuv_mstar}
\end{figure}

Figure~\ref{fig:afuv_mstar} shows the relation between stellar mass
$M_*$ and dust attenuation in terms of $\ltir/\lfuv$.  At a first
glance, the dependence of dust attenuation on stellar mass is very
strong.  Though we should be cautious on the selection effect on this
plot again, the strong dependence might be partially physical.  This
implies that larger galaxies are more extinguished.  Hence, dust
attenuation is closely related to the physical size of galaxies.

The presence of low mass and high $\ltir/\lfuv$ galaxies $\sim
10\mbox{--}100$ at $M_* \simeq 10^7\mbox{--}10^9\;M_\odot$ is worth
mentioning here.  \citet{iglesias06} made essentially the same
analysis by GALEX and IRAS.  In their studies, such galaxies did not
exist in their IR-selected samples (their Figure~12).  If the flux
measurement is secure, these galaxies deserve a close inspection to
examine their nature.

\begin{figure}[t]
\centering\includegraphics[width=0.4\textwidth]{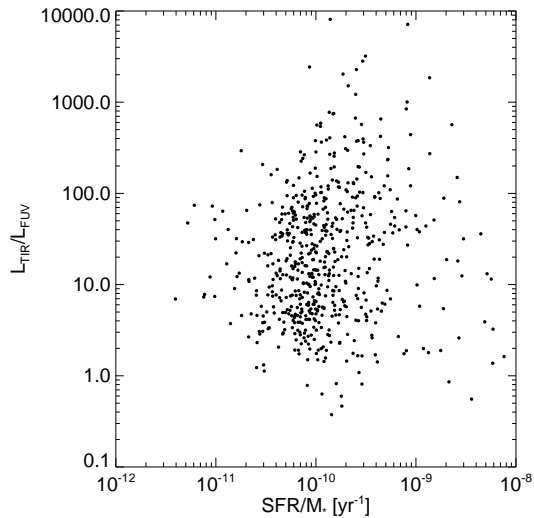}
\caption{
Relation between SFR per unit stellar mass, $\mbox{SFR}/M_*$ (specific SFR: SSFR),
{and dust attenuation $\ltir/\lfuv$.
}
}\label{fig:afuv_ssfr}
\end{figure}

Next, we show the relation between SFR per unit stellar mass,
$\mbox{SFR}/M_*$, which is often referred to as the specific SFR
(SSFR), {and dust attenuation $\ltir/\lfuv$} in
Figure~\ref{fig:afuv_ssfr}.  It is indeed striking that obviously
there is no correlation between these quantities.

First conclusion would be that there is no correlation
between global attenuation and SSFR.
Generally, many authors showed that
dust attenuation increases with SFR \citep[see, e.g., Figure~7
of][]{buat07a}.  Then, if there is no link with the SSFR it implies
that dust attenuation is linked to the total amount of SF scaled with
galaxy size, because SFR was positively correlated to $M_*$ seen
above.  It may be interpreted as follows: the attenuation is not
related to the current-to-past SFR ratio, or roughly, the SF history.
Probably this could be explained by the short lifetime of dust grains
in the ISM.

We should mention that, however, \citet{iglesias06} has shown that
even for the FIR-selected galaxies there is some correlation between
these two quantities (see their Figure~10b).  This issue seems to
require further examination.

\section{Conclusion}\label{sec:conclusion}

In order to explore SEDs, star formation, and dust extinction
properties of galaxies in the Local Universe, we have constructed a
multiband galaxy sample based on the AKARI FIS All-Sky Survey
and GALEX All-Sky Imaging Survey (AIS).  
We start from AKARI All-Sky
Survey Bright Source Catalogue $\beta$-1.1, and selected galaxies by
matching the AKARI sources with those of the IRAS PSC$z$.  
Next, we have measured total GALEX FUV and NUV flux densities by a photometry
software which we have developed specifically for this purpose.  
Then, we have matched this sample with
SDSS and 2MASS galaxies to obtain the basic sample.  
The basic sample consists of 776 galaxies.  
{
After removing objects with photometry contaminated by foregound 
sources (mainly in SDSS), we have defined the ``secure sample'' which
contains 607 galaxies. 
}
Based on this
galaxy sample, we have explored various properties of galaxies related
to star formation and dust extinction.

Summary and conclusions of this study are as follows:
\begin{enumerate}
\item {The sample galaxies have redshifts $\la 0.15$}, and
	their 90-$\mu$m luminosities range from $10^6$ to  $10^{12}\;L_\odot$, 
	with a peak at $10^{10}\;L_\odot$.
\item {The SEDs display a very large variety, especially more than four 
	orders of magnitude at M-FIR}, but if we sort the sample by $90\;\mu$m, 
	their average SED has a coherent trend: the more
	luminous at $90\;\mu$m, the redder the global SED becomes.
\item The $M_r$-$\mbox{NUV}-r$ color-magnitude relation of our sample
	does not show a bimodality which is almost always expected in
	optically selected galaxy samples.
	The distribution is unimodal, centered on the green valley between blue
	cloud and red sequence seen in optical surveys.
\item We have established formulae to convert FIR luminosity from
	AKARI bands to the total infrared (IR) luminosity $\ltir$.
\item The luminosity related to star formation activity ($\lsf$) is
	dominated by the contribution of $\ltir$ even if we take into
	account the FIR emission from dust heated by old stars.
\item With these formulae, we calculated the star formation directly
	visible with FUV and hidden by dust. 
	At high star formation rate (SFR) ($> 20\;\mbox{M}_\odot\,\mbox{yr}^{-1}$), 
	the fraction of directly visible SFR, $\sfr_{\rm FUV}$, decreases.
\item We estimated the ratio, $\ltir/\lfuv$, which is a direct measure
	of the FUV attenuation $A_{\rm FUV}$.  
	The distribution of
	$\ltir/\lfuv$ is consistent with a previous result based on GALEX
	and IRAS \citep{buat07a}.
\item We also examined the $\ltir/\lfuv$-UV slope ($\mbox{FUV}-
	\mbox{NUV}$) relation.  
	The majority of the sample
    has $\ltir/\lfuv$ ratios 5 to 10 times lower than expected from
    the local starburst relation \citep{kong04}, while some LIRGs and
    all the ULIRGs of this sample have higher $\ltir/\lfuv$ ratios.
	This trend was already reported from a previous GALEX-IRAS study
	\citep{buat05} obtained by a much smaller sample, and we have
	confirmed their conclusion.
\item By making use of stellar mass information derived from SDSS flux densities 
	in this work), we have examined the dust attenuation properties in terms of stellar
	mass.  
	We found that the attenuation indicator $\ltir/\lfuv$ is
	correlated to stellar mass of galaxies, $M_*$, but there is no
	correlation with specific SFR (SSFR), $\mbox{SFR}/M_*$.  
	This may mean that $\ltir/\lfuv$ is not linked to the SF history, 
	but simply scales with the size of galaxies.  
	However, this is at odds with previous result of \citet{iglesias06}.
\end{enumerate}

This sample will serve as an important reference sample at $z=0$ for
various further analysis or ongoing/future observational projects,
like Herschel: for instance, this can be used to construct a set of
SEDs for discussing higher-$z$ observational strategy, or as a
baseline test sample to investigate a method of extracting galaxies
only from FIR flux information \citep[e.g.,][]{pollo10}.

However, since our first sample is not complete in many senses,
further analysis will be desired.  
We plan to construct a larger
multiwavelength sample from the next release of AKARI FIS All-Sky
Survey in the near future.

\begin{acknowledgements}
We deeply thank the anonymous referee for her/his careful
reading of the original manuscript, useful
suggestions and comments which improved the clarity of the paper.
This work is based on observations with AKARI, a JAXA
project with the participation of ESA.  
TTT has been supported by
Program for Improvement of Research Environment for Young
Researchers from Special Coordination Funds for Promoting Science
and Technology, and the Grant-in-Aid for the Scientific Research
Fund (20740105) commissioned by the Ministry of Education, Culture,
Sports, Science and Technology (MEXT) of Japan.  
VB and DB have been
supported by the Centre National des Etudes Spatiales (CNES) and the
Programme National Galaxies (PNG).  
We thank Agnieszka Pollo, Mai
Fujiwara, Akira Ikeyama, Ryosuke Asano, Akio K.\ Inoue, Hiroshi
Shibai, Yasuo Doi, Hideaki Fujiwara, Mitsunobu Kawada,
Hidehiro Kaneda, Hiroyuki Hirashita, and TakakoT.\ Ishii for
fruitful discussions and comments.  
TTT, FTY, and KLM are partially
supported from the Grand-in-Aid for the Global COE Program ``Quest
for Fundamental Principles in the Universe: from Particles to the
Solar System and the Cosmos'' from the MEXT.
\end{acknowledgements}

\appendix

\section{Luminosity vs. redshift diagrams of the sample}\label{sec:zlum_all}

\begin{figure*}[t]
\centering\includegraphics[width=0.24\textwidth]{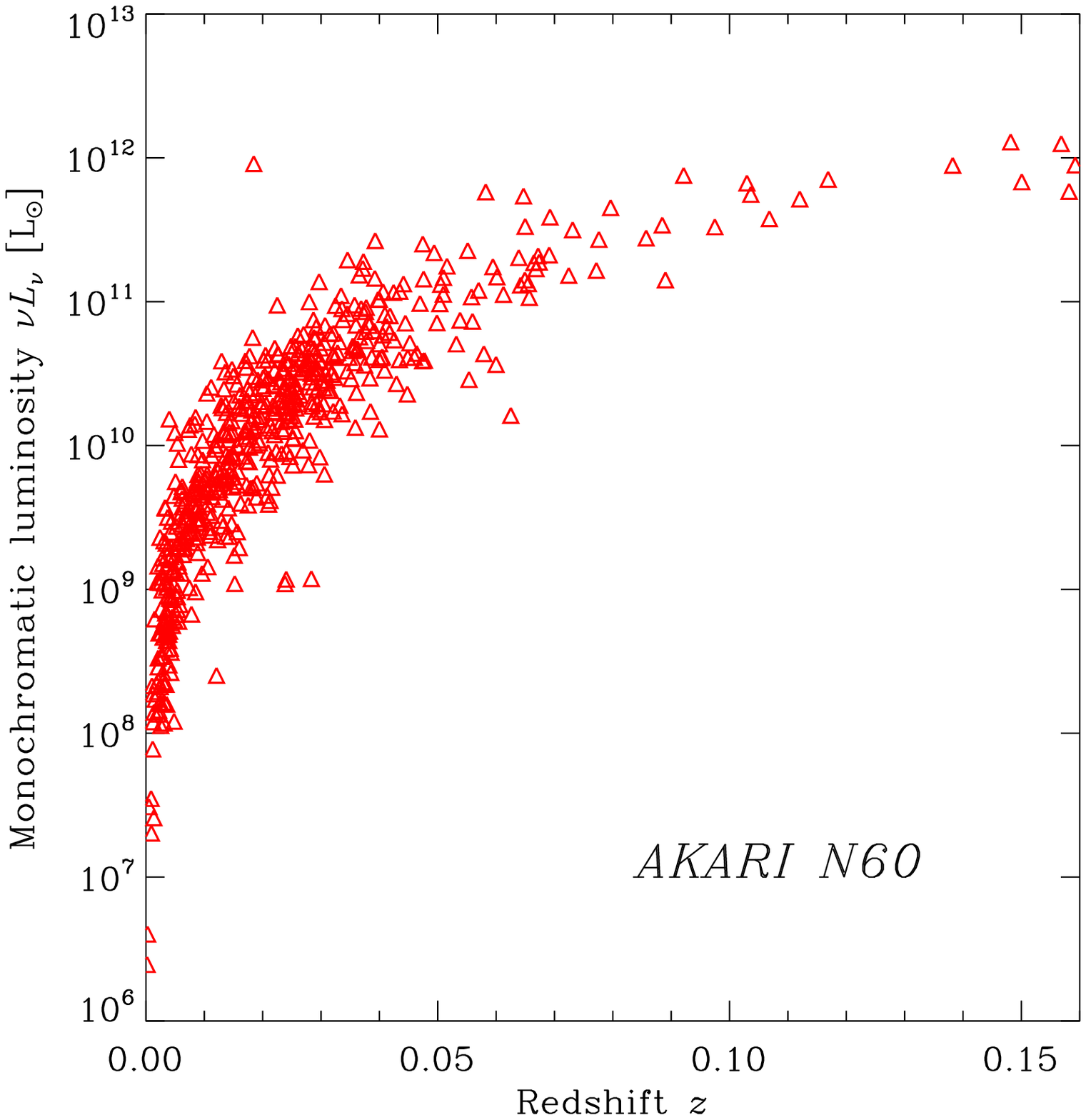}
\centering\includegraphics[width=0.24\textwidth]{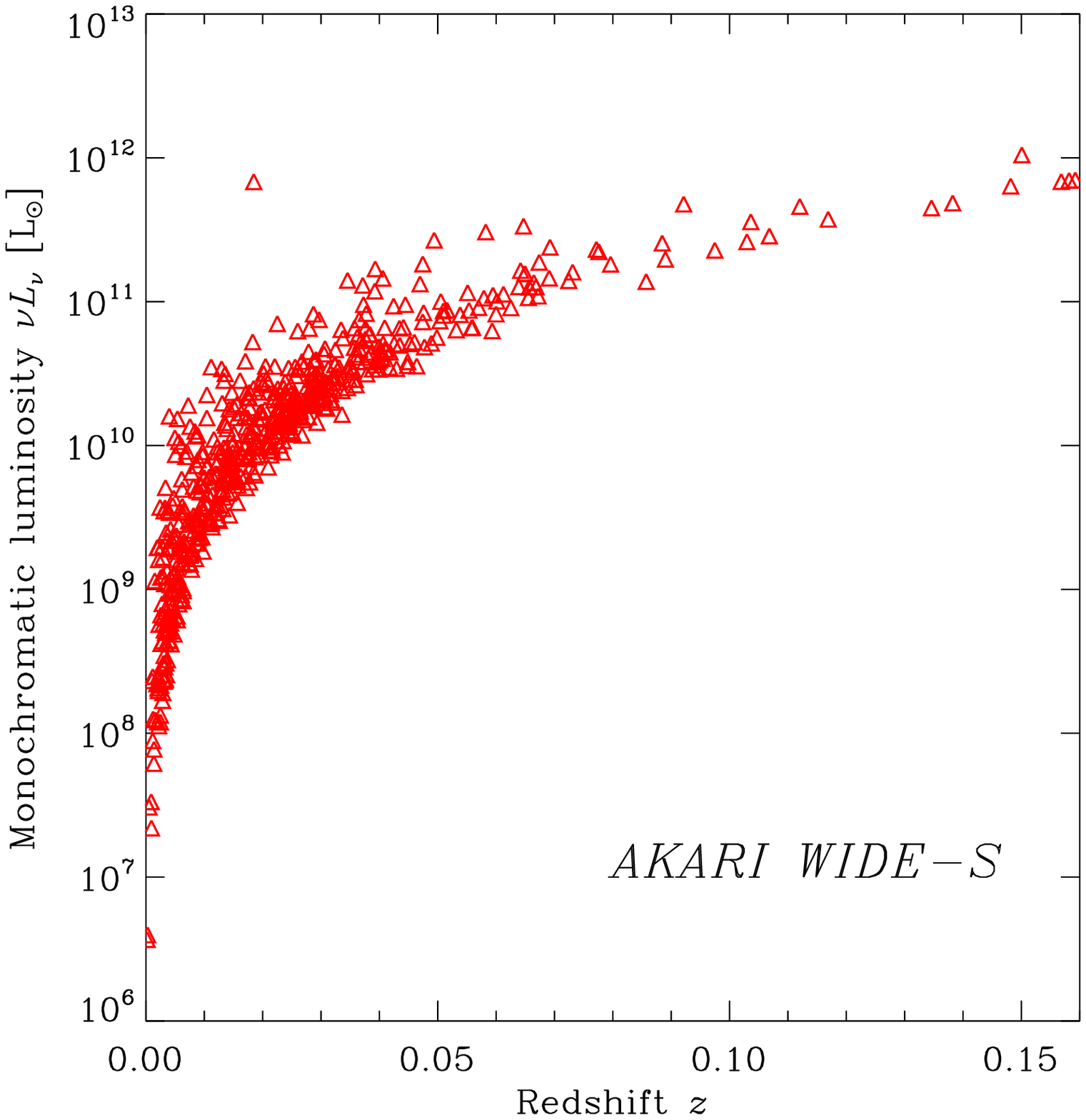}
\centering\includegraphics[width=0.24\textwidth]{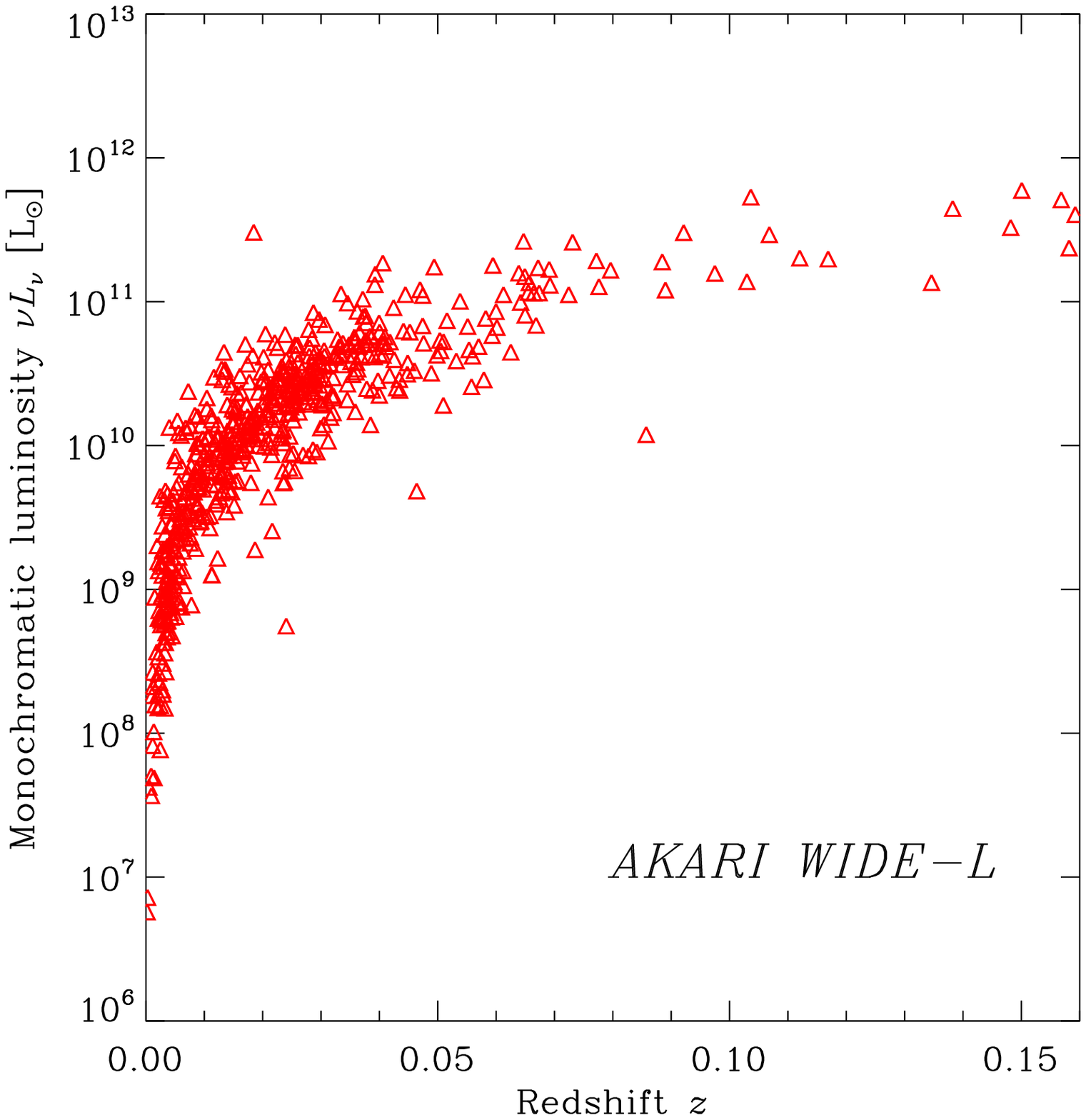}
\centering\includegraphics[width=0.24\textwidth]{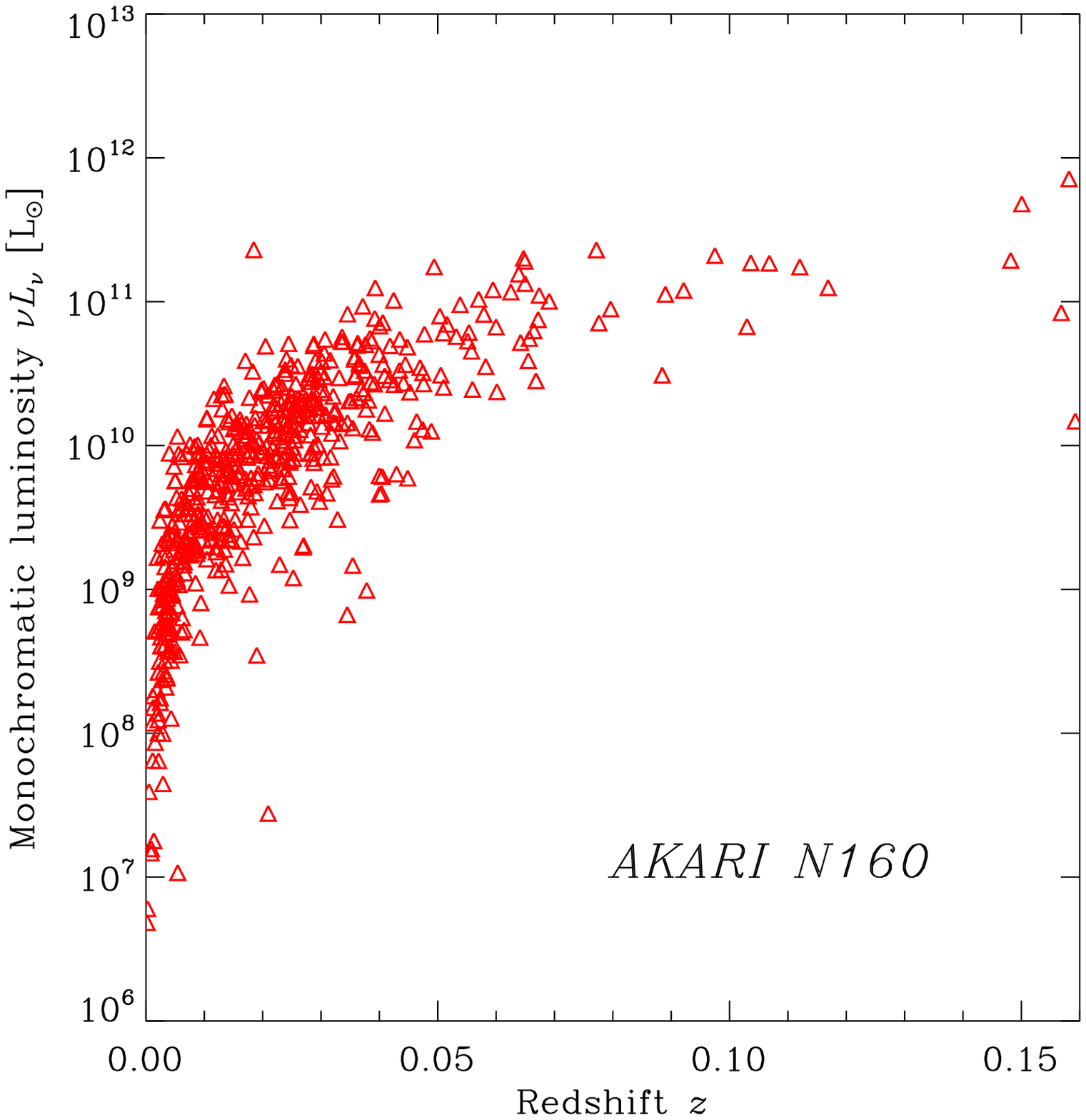}
\centering\includegraphics[width=0.24\textwidth]{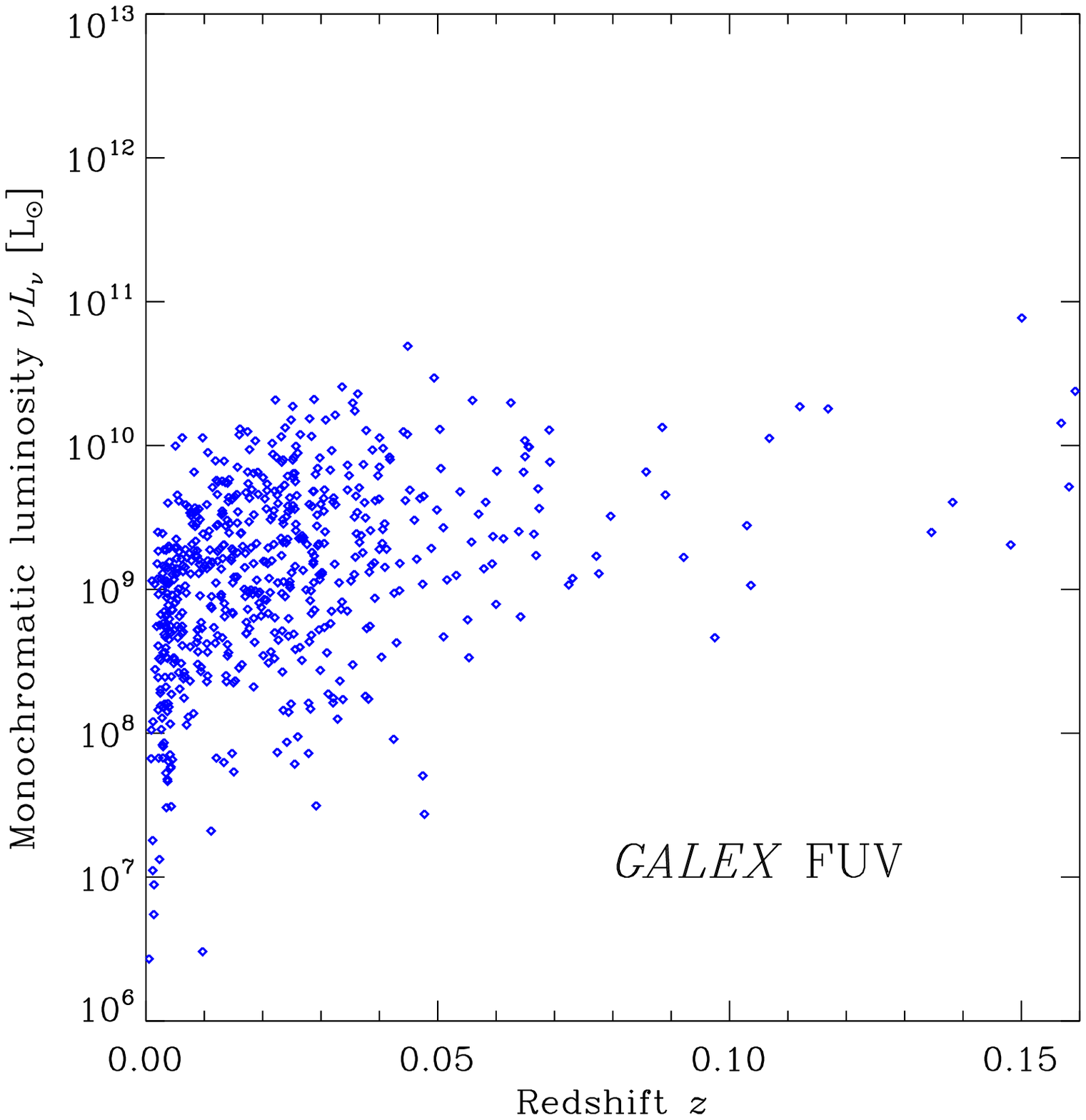}
\centering\includegraphics[width=0.24\textwidth]{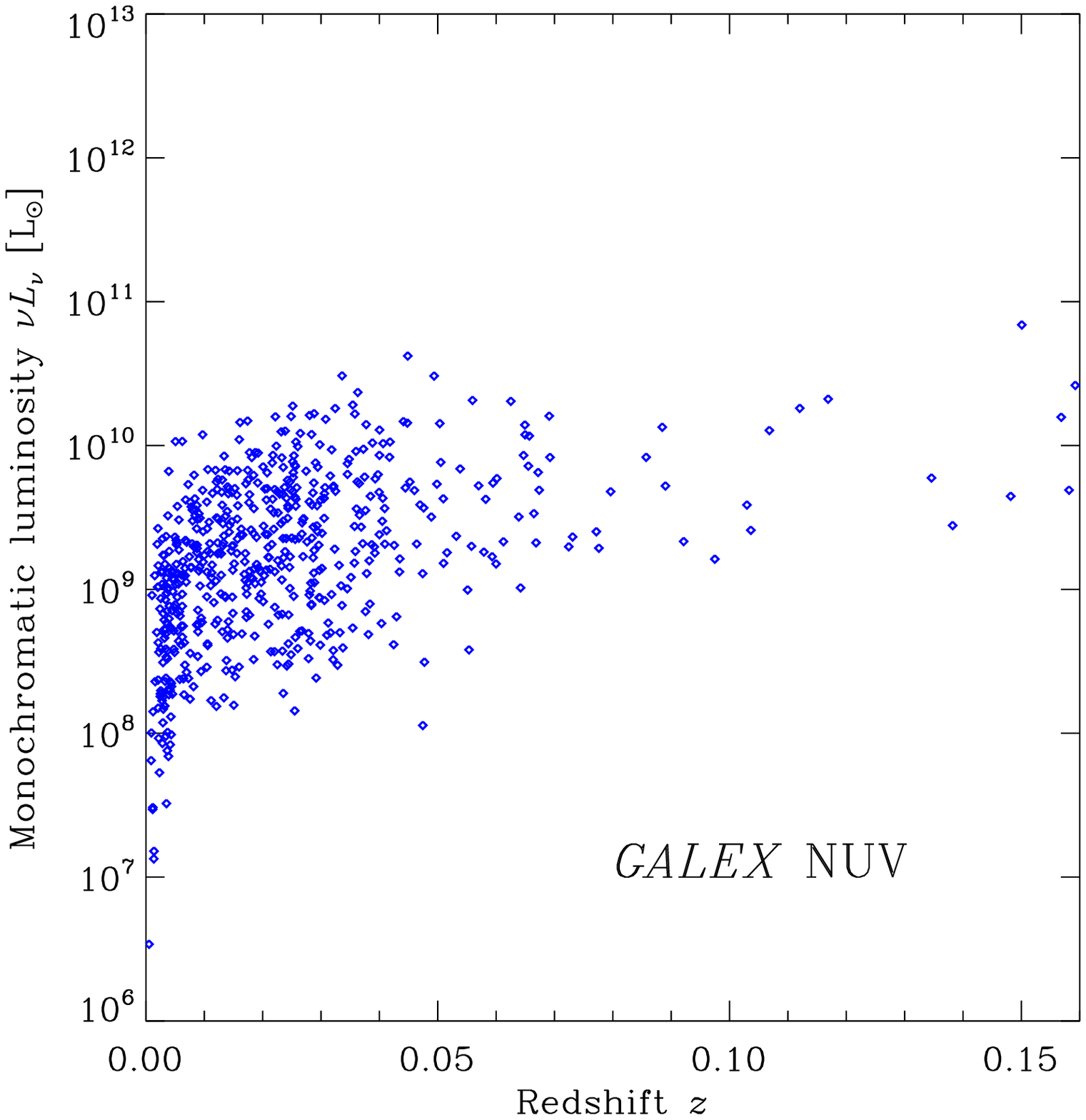}
\centering\includegraphics[width=0.24\textwidth]{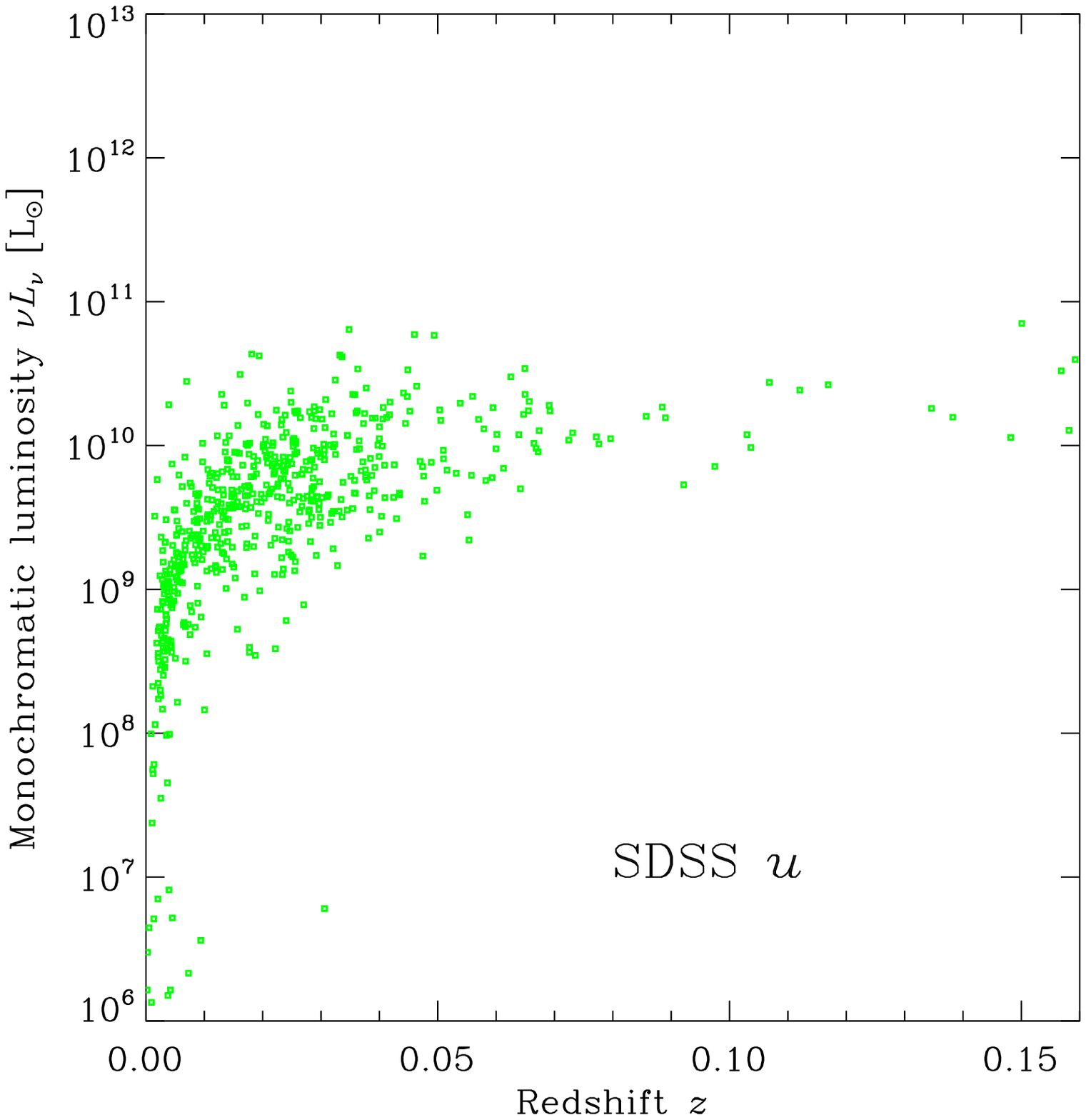}
\centering\includegraphics[width=0.24\textwidth]{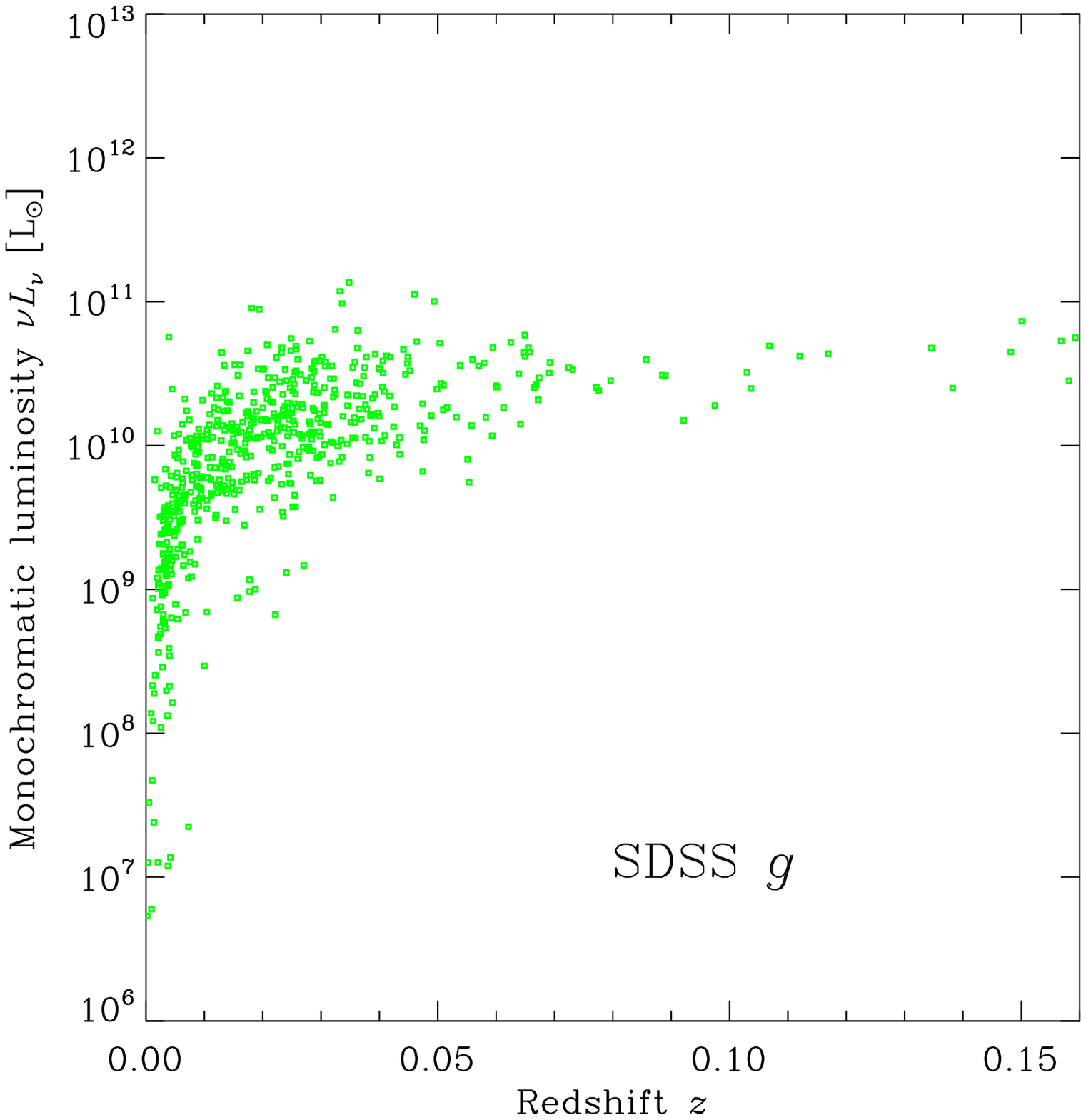}
\centering\includegraphics[width=0.24\textwidth]{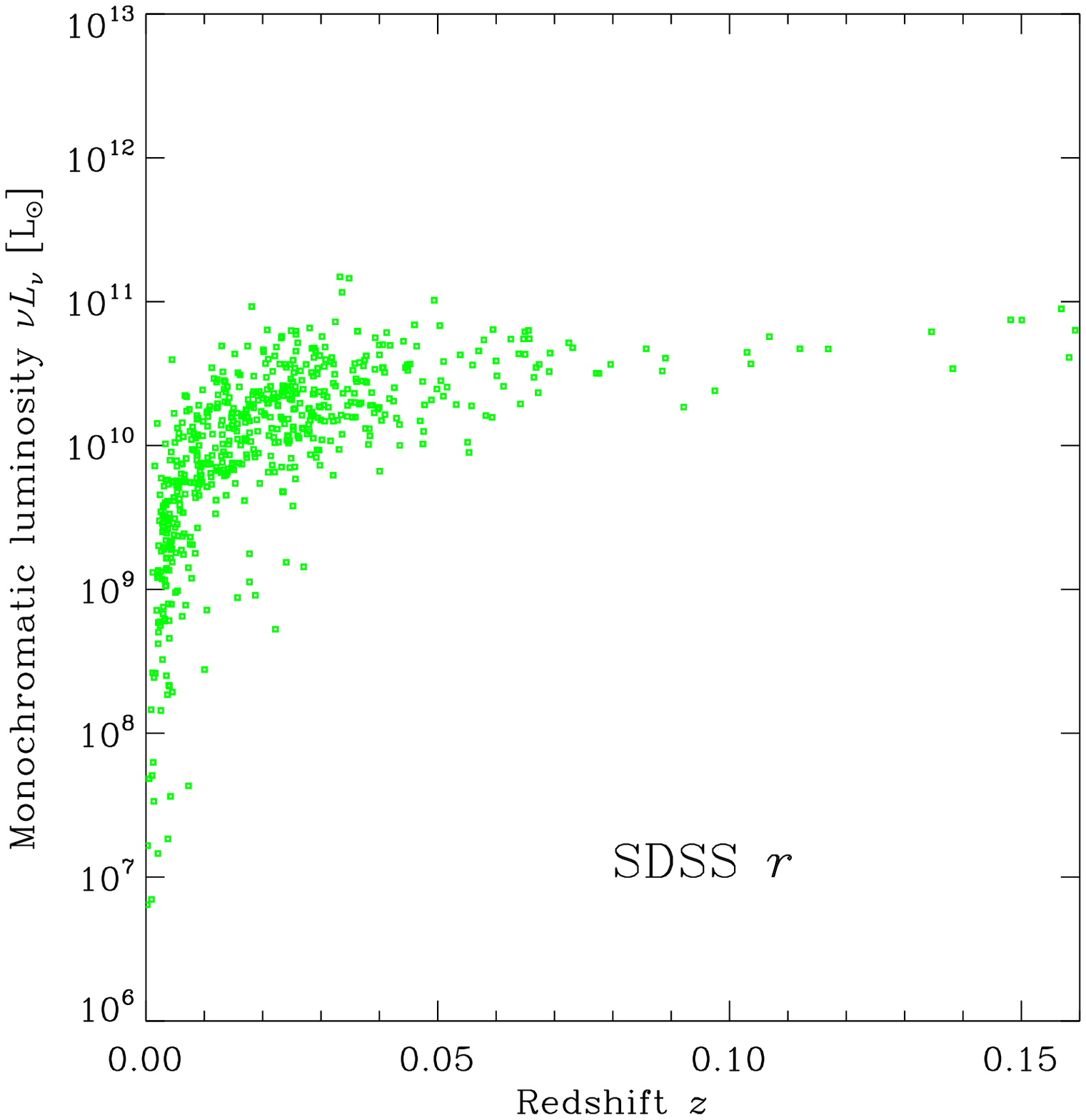}
\centering\includegraphics[width=0.24\textwidth]{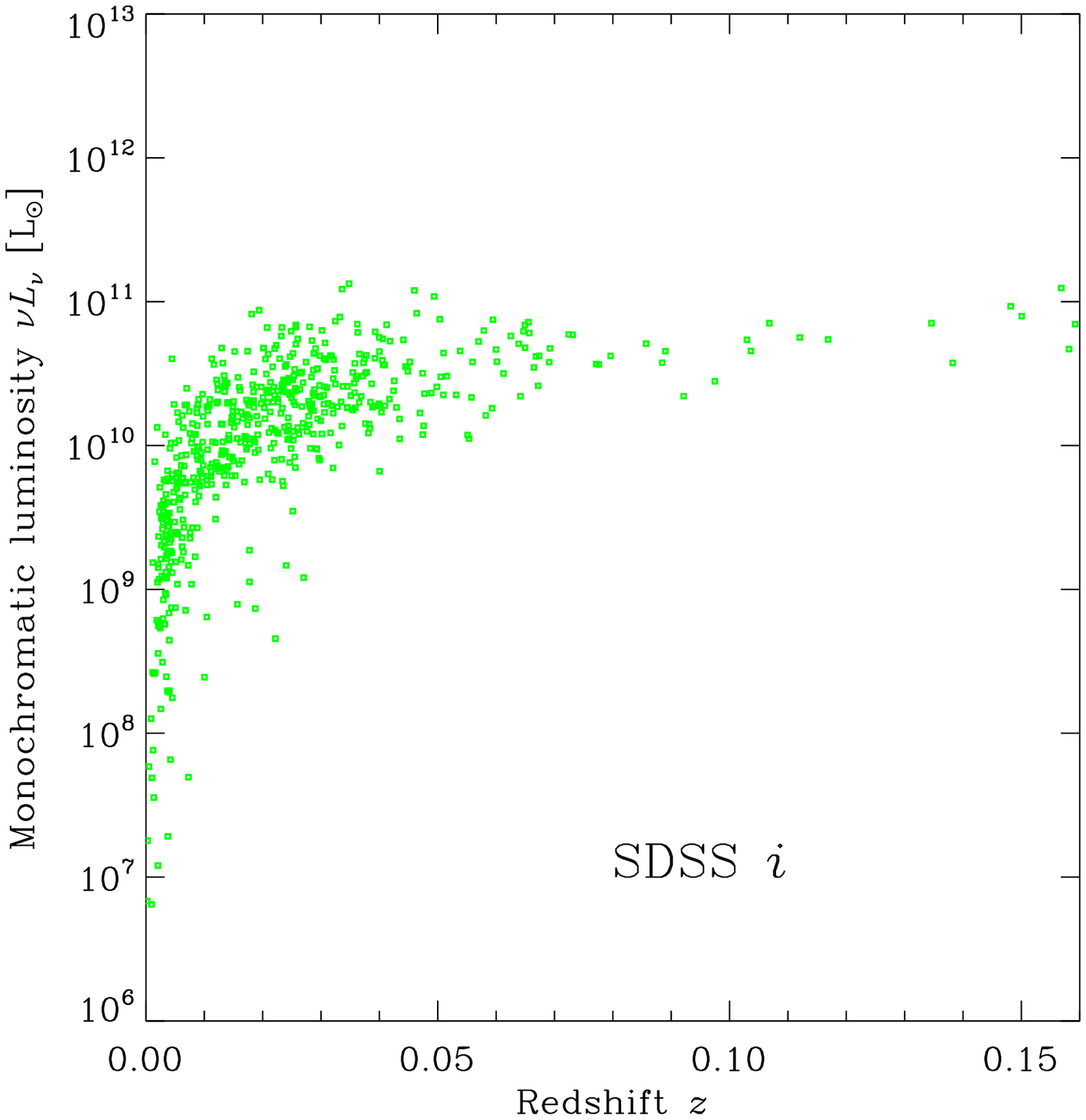}
\centering\includegraphics[width=0.24\textwidth]{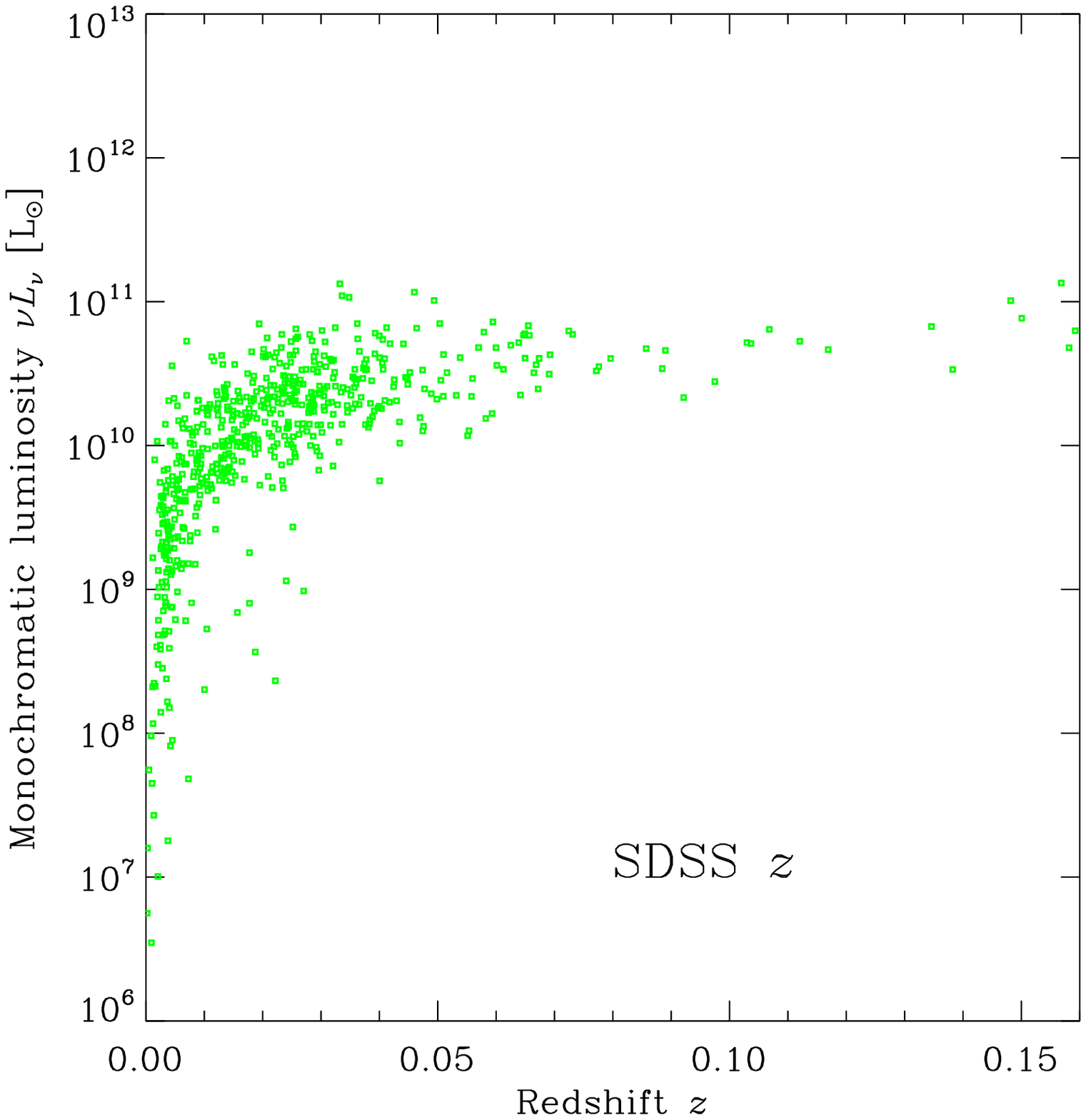}
\centering\includegraphics[width=0.24\textwidth]{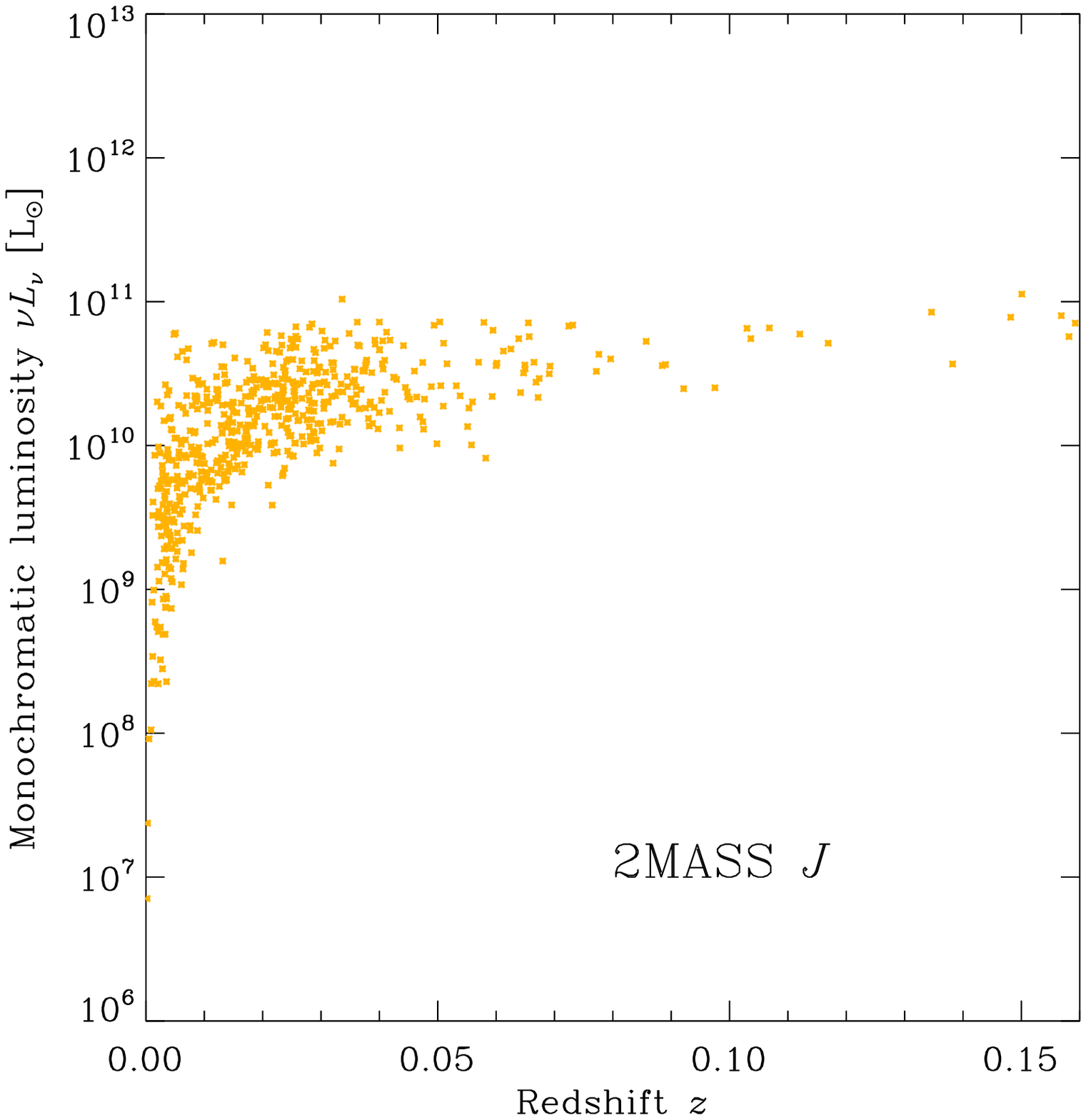}
\centering\includegraphics[width=0.24\textwidth]{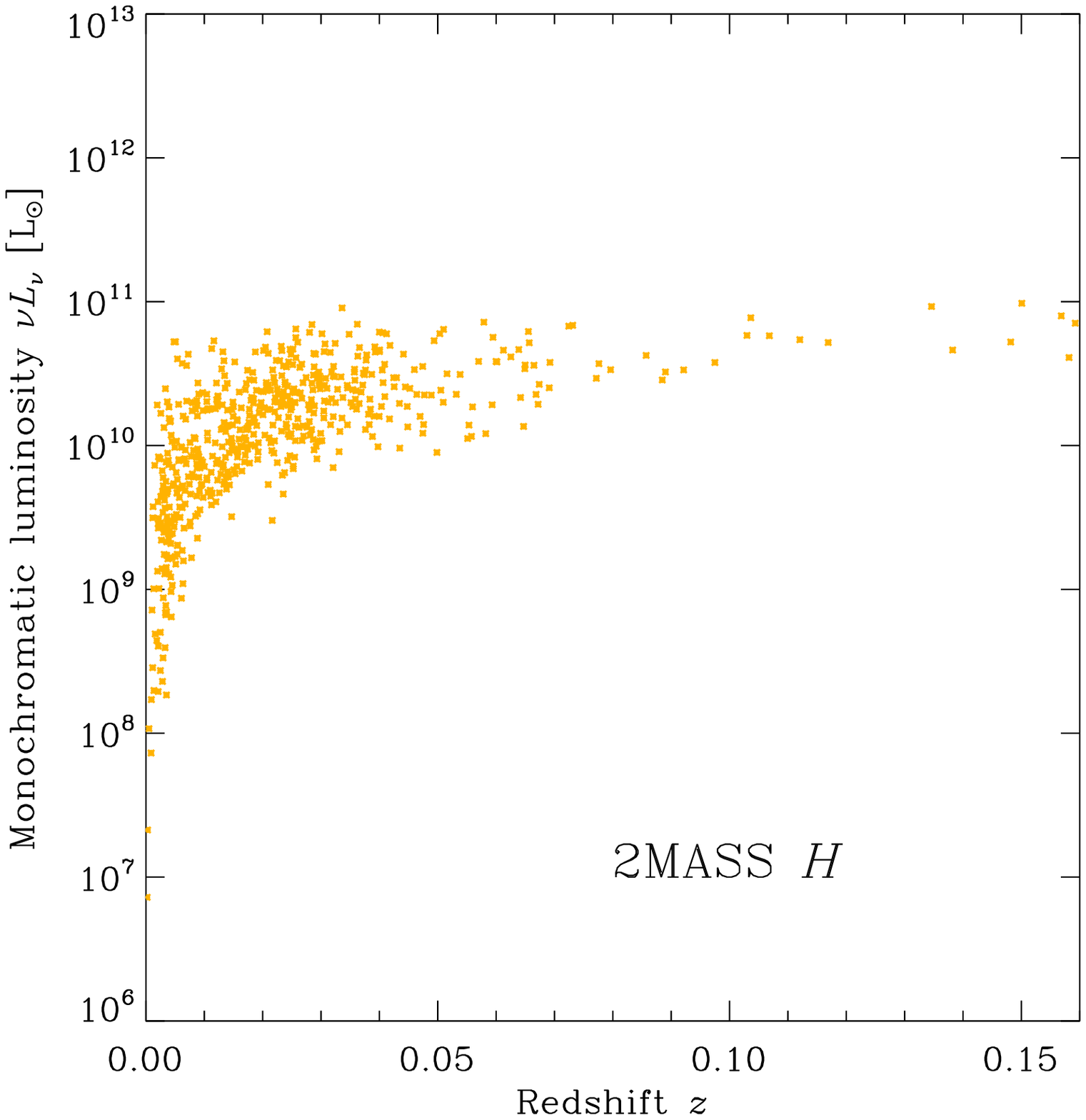}
\centering\includegraphics[width=0.24\textwidth]{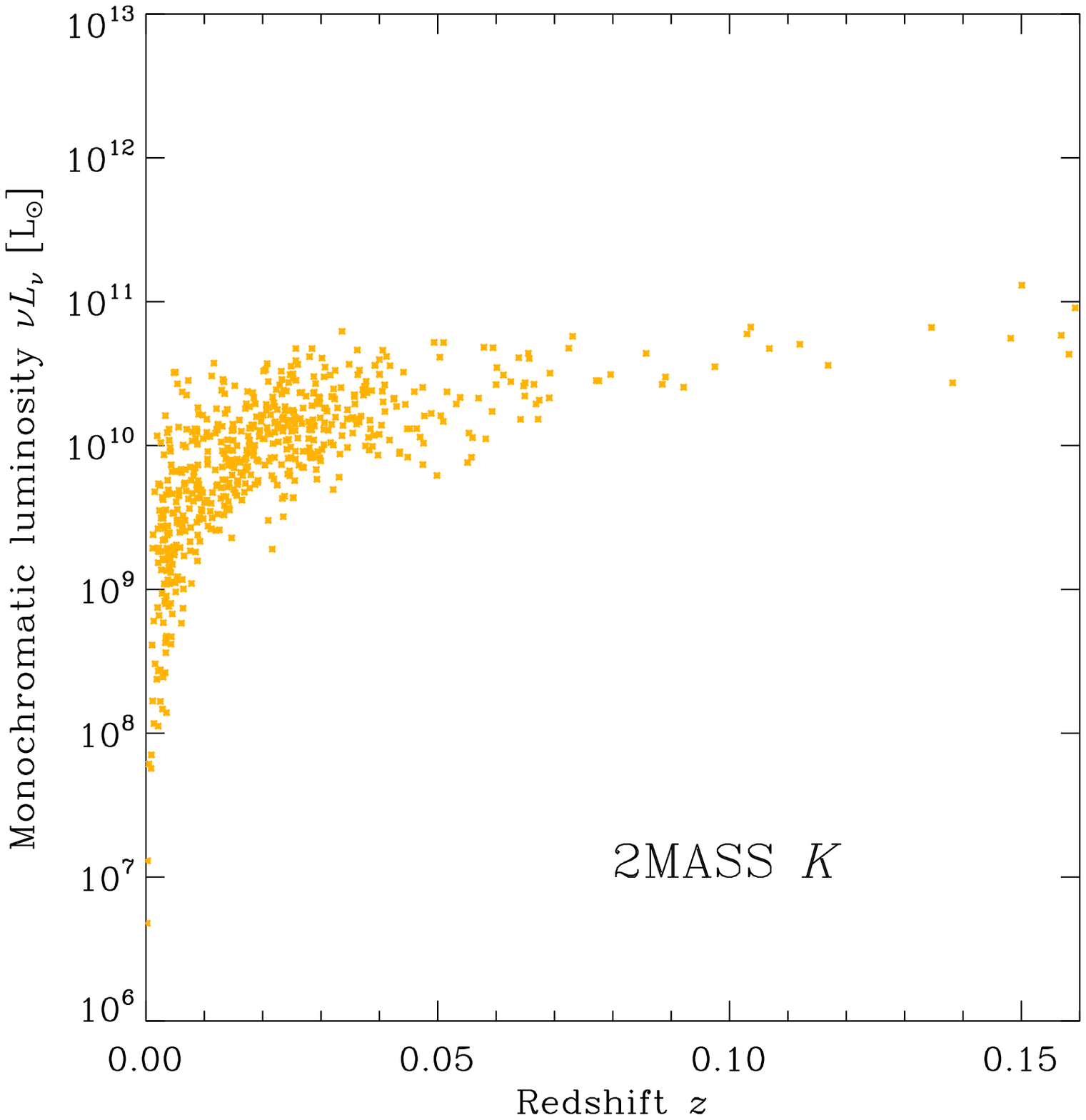}
\centering\includegraphics[width=0.24\textwidth]{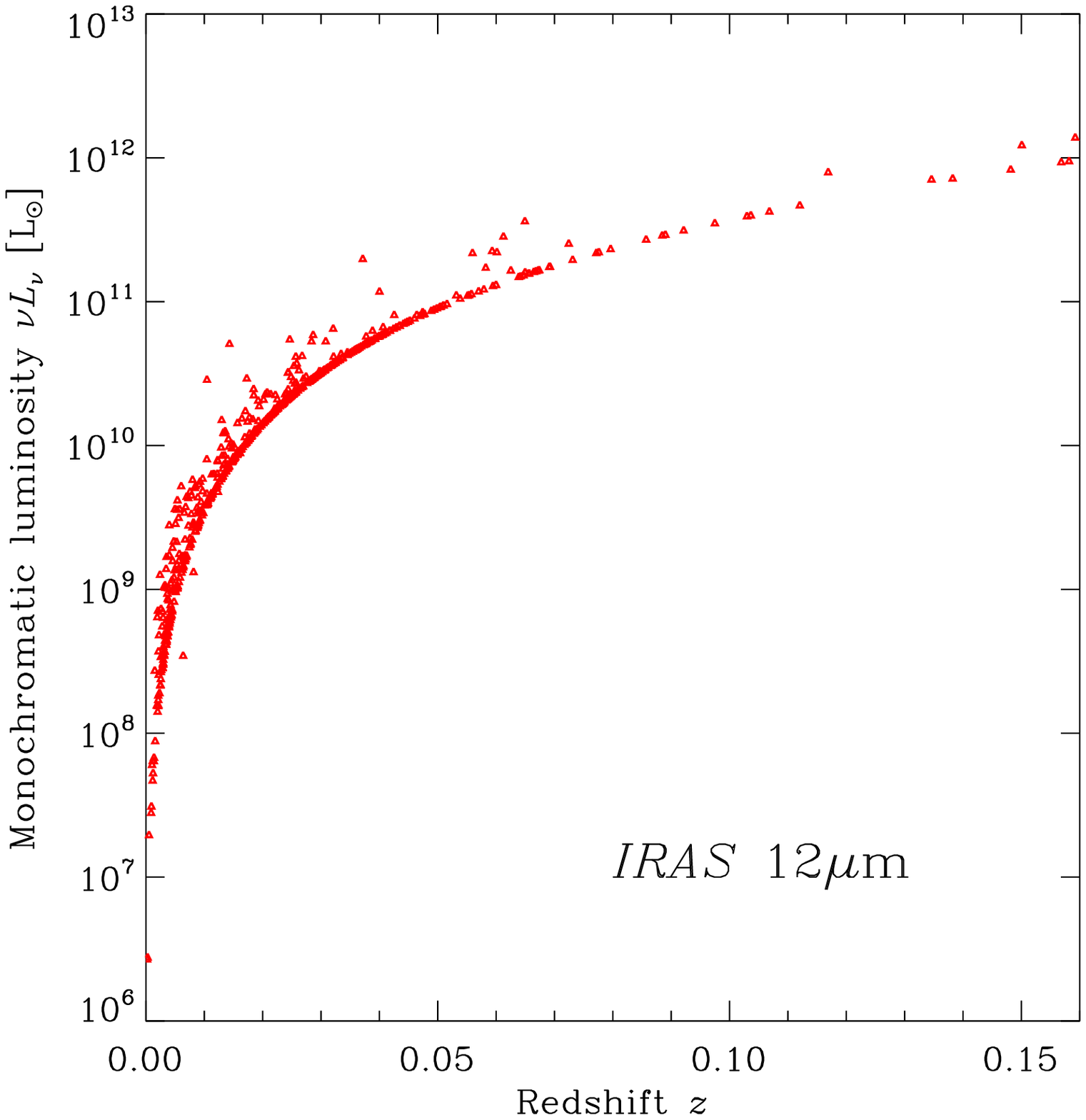}
\centering\includegraphics[width=0.24\textwidth]{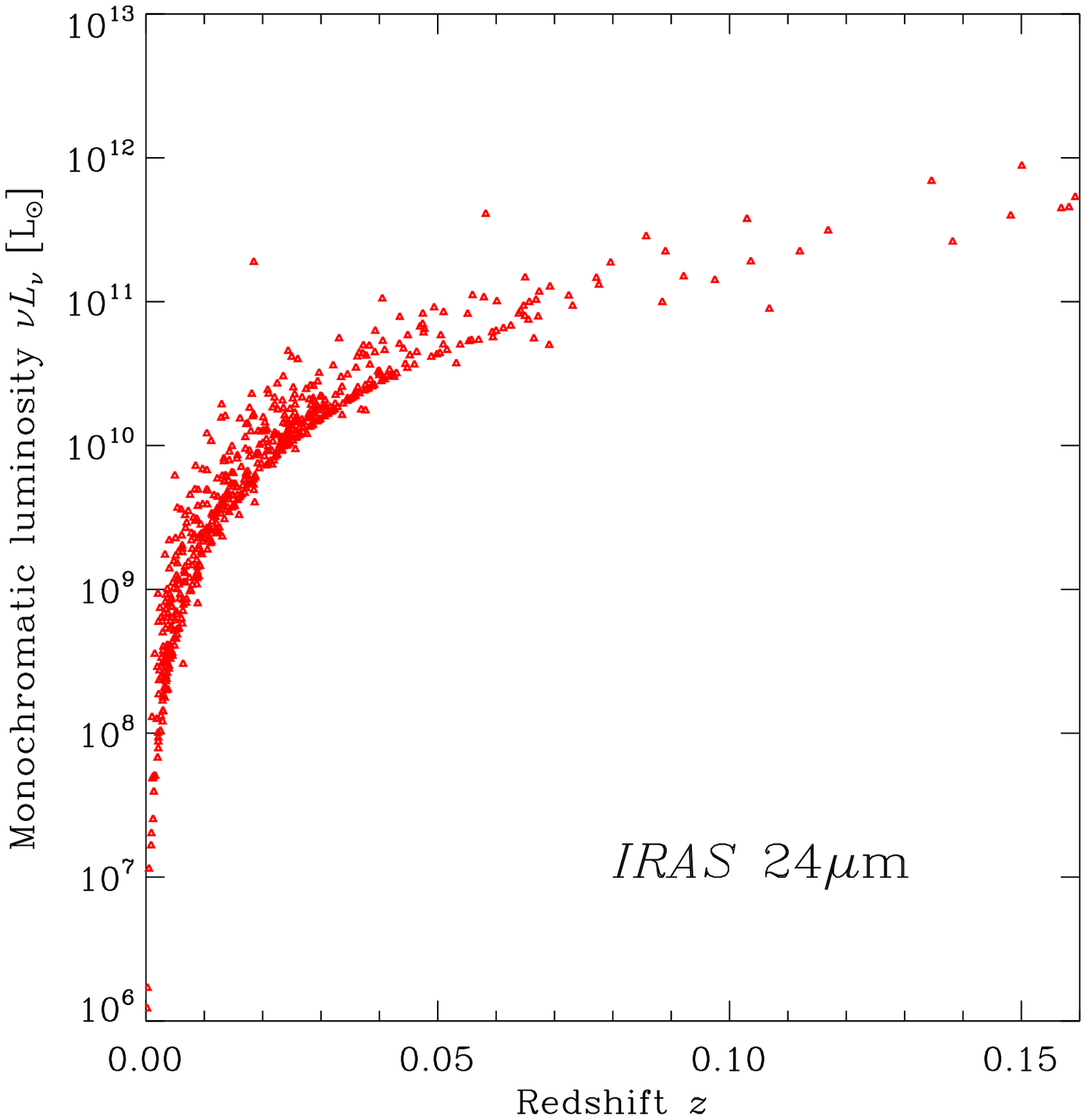}
\centering\includegraphics[width=0.24\textwidth]{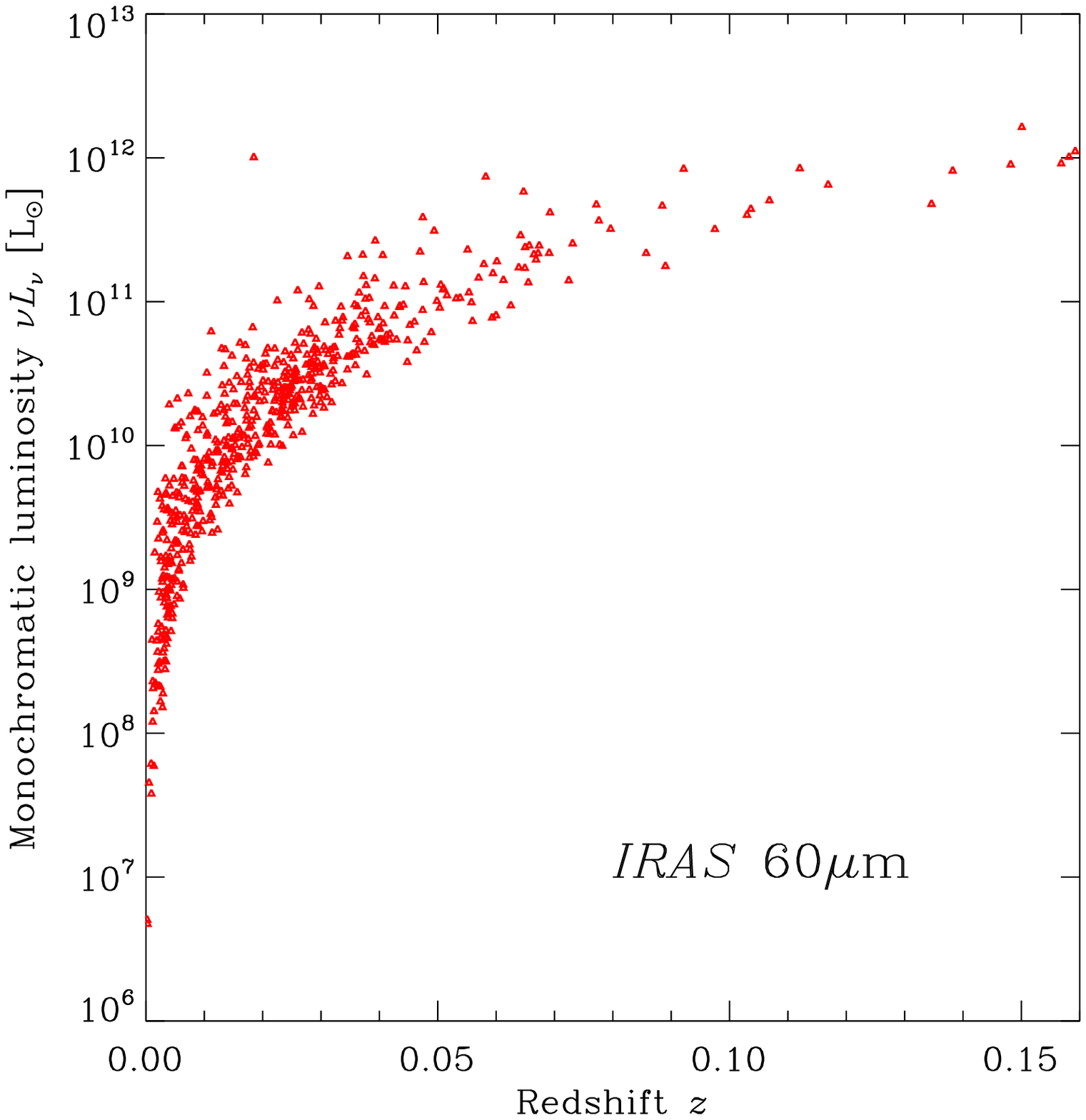}
\centering\includegraphics[width=0.24\textwidth]{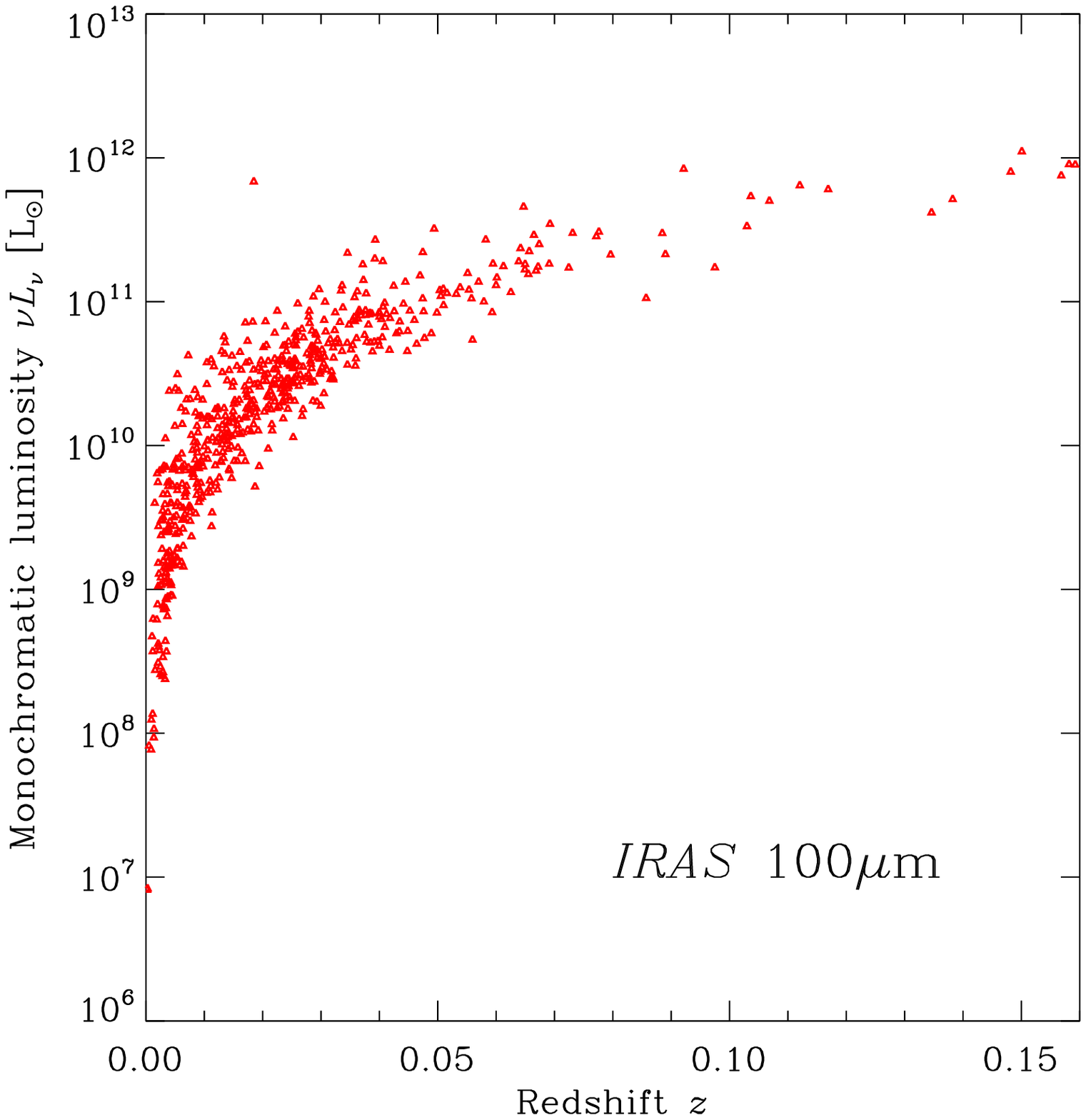}
\caption{Luminosity vs. redshift diagrams for all the bands in the sample.}\label{fig:lum}
\end{figure*}

In Appendix, we show the redshift--luminosity relations for all the bands we have
used in the main text (Fig.~\ref{fig:lum}).
These plots help to examine the effect of the selection functions of our sample
at each band.

\end{document}